\documentclass{jpp}
\usepackage{graphicx}
\usepackage{hyperref}

\usepackage[utf8]{inputenc}
\usepackage[T1]{fontenc}
\usepackage{amsmath}
\usepackage{colortbl}

\shorttitle{Near-axis description of quasi-isodynamic stellarators to second order}
\shortauthor{E. Rodriguez, G. G. Plunk
, R. Jorge
}

\title{Near-axis description of stellarator-symmetric quasi-isodynamic stellarators to second order}

\author{E. Rodríguez\aff{1}, G. G. Plunk\aff{1}
, R. Jorge\aff{2}
}

\affiliation{
\aff{1} Max Planck Institute for Plasma Physics, 17491 Greifswald, Germany
\aff{2} Department of Physics, University of Wisconsin-Madison, Madison, WI, USA 
}

\begin{document}

\maketitle

\begin{abstract}
    The near-axis description of optimised stellarators, at second order in the expansion, provides important information about the field, both of physical and practical importance for stellarator optimisation.  It however remains relatively underdeveloped for an important class of such stellarators, called quasi-isodynamic (QI).  In this paper we develop the theoretical and numerical framework for the construction of such solutions.  We find that the case of QI stellarators calls for the careful treatment of continuity, smoothness and periodicity of the various functions involved, especially for so-called half-helicity fields, which feature prominently in existing QI designs.  The numerical implementation of necessary elements is described, and several examples are constructed and quantitatively verified in detail. This work establishes a basis for further systematic exploration of the space of QI stellarators, and the development of both theoretical and practical tools to facilitate effective optimisation of QI stellarators. 
\end{abstract}

\section{Introduction}
Understanding and designing stellarators \citep{spitzer1958stellarator,boozer1998stellarator,helander2014theory} can be a daunting task when one considers how large the space of general toroidally shaped three-dimensional magnetic fields is. Naturally, the vast majority of such fields are uninteresting, as most will fail to confine plasmas in a finite volume for long enough to undergo thermonuclear fusion, which is the main reason for their study. 
\par
The requirement of confinement is satisfied in a certain sense for an interesting subclass of stellarators: \textit{omnigeneous stellarators} \citep{bernardin1986,Cary1997,hall1975,helander2014theory}. Omnigeneous fields are fields that, by definition, confine all collisionless charged particle orbits \citep{northrop1961guiding,littlejohn1983,wessonTok,blank2004guiding}, and thus are optimised in that regard. To achieve such a behaviour, fields must present a carefully tailored magnetic field magnitude $|\mathbf{B}|$ \citep{boozer1983transport,nuhren1988,Cary1997,parra2015less}, which is coupled in a rather complex way to the geometry of the field. This requires careful optimisation \citep{mynick2006}.
\par
Understanding of these fields, as well as a practical procedure to provide initial seeds for large scale optimisation, require a more controlled, simplified perspective on the problem. One such perspective has been historically provided by a \textit{near-axis description} of the field \citep{mercier1964equilibrium,Solovev1970,lortz1976equilibrium,garrenboozer1991a}: an asymptotic description of the equilibrium field in the distance from its centre (called the magnetic axis). In such a context, the geometry and field descriptions simplify significantly, proving a powerful tool in advancing the theoretical understanding of optimised omnigeneous stellarators \citep{mercier1964equilibrium,lortz1976equilibrium,landreman2020magnetic,landreman2021a,jorge2020naeturb,rodriguez2022phases,rodriguez2024maximum,rodriguez2023mhd} as well as providing a practical tool in stellarator design \citep{landreman2019,landreman2022mapping,rodriguez2023constructing,jorge2022c,camacho-mata-2022}. This framework has reached a certain level of maturity specially within a particular subclass of optimised stellarators: namely \textit{quasisymmetric} stellarators \citep{boozer1983transport,nuhren1988,rodriguez2020necessary,burby2020some}. These fields are characterised by a direction of symmetry in $|\mathbf{B}|$ either in a toroidal (QA) or helical (QH) direction, a symmetry that simplifies their description in a way that does not occur in the broader sense of omnigeneous fields. This symmetry difference has proven to make it theoretically challenging to describe the other big class of omnigeneous fields, so called \textit{quasi-isodynamic} (QI) fields \citep{Cary1997,Helander_2009,Nührenberg_2010}, which have poloidally closed $|\mathbf{B}|$-contours, but although omnigeneous, not a direction of symmetry. As a result, the near-axis description of QI fields to date has been restricted to its most reduced (first order) form, i.e. elliptically shaped cross-sections, with no information regarding key properties such as MHD stability or triangularity \citep{plunk2019direct,jorge2022c,camacho-mata-2022,Camacho2023helicity}. Recently, the omnigeneity conditions required by QI at higher order have been presented \citep{rodriguez2023higher}, but the work did not go as far as to attempt a fully consistent treatment including the solution of the equilibrium equations.
\par
In this paper we set ourselves the task of bringing this near-axis framework suited to stellarator symmetric \citep{dewar1998stellarator} QI stellarators en par of the quasisymmetric one. We do so by appropriately extending the considerations of equilibrium and omnigeneity (and their cohabitation) to second order in the near-axis expansion. In Section~\ref{sec:nae_equilibrium} we present the near-axis description of a stellarator-symmetric equilibrium field with poloidal $|\mathbf{B}|$ contours, focusing on providing a clear physical picture for the set of equations involved, originally presented in \cite{landreman2019}. Special emphasis is placed on those aspects that are a consequence of the poloidal topology of $|\mathbf{B}|$, and thus distinguish this case from the quasisymmetric one. The interaction of the equilibrium with omnigeneity is the focus of the next section, Section~\ref{sec:2nd_order_qi_conditions}. Finally, we present a number of numerical examples in which the near axis constructions are compared to global equilibria, providing a benchmark of the near-axis construction to second order, setting the ground to using this framework for future applications.

\section{Near-axis equilibria to second order} \label{sec:nae_equilibrium}
The near-axis expansion is an asymptotic description of the magnetic field and its properties near a central closed magnetic field line, which is called the \textit{magnetic axis} \citep{mercier1962,Solovev1970,lortz1976equilibrium,garrenboozer1991a}. The power of the description rests on the simplicity of the fields near this closed curve, assumed to form nested flux surfaces. 
\par
In this section we set-up the near-axis description of an equilibrium magnetic field, assuming poloidally closed contours of field strength. This is a necessary condition for a QI stellarator, although not sufficient. The description is completed in the next section, where omnigeneity is introduced. The general set of equations governing this description and how to algebraically obtain them is presented in detail in the work of \cite{landreman2019} (henceforth `LS'), in particular Appendix~A therein.
\par
\subsection{General problem set-up}
Let us set up the equilibrium problem by introducing the governing set of equations that the magnetic field $\mathbf{B}$ must satisfy. First, for the field to represent a magnetic field, it must be solenoidal, i.e. (i) $\nabla\cdot\mathbf{B}=0$. In addition, we shall consider the field to have nested toroidal surfaces labelled by $\psi$ (the toroidal flux enclosed by the flux surfaces over $2\pi$) tangent everywhere to the field. That is, (ii) $\mathbf{B}\cdot\nabla\psi=0$. An appropriate degree of smoothness of this toroidal folliation of space is assumed (especially in the neighbourhood of the field axis, where the near-axis expansion will ensue) \citep{burby_duignan_meiss_2021_NAE_coordinates,Duignan_Meiss2021normal_form}. Finally, we must impose the equilibrium condition, which we do in its simplest form (the static limit of MHD \citep{Kruskal-Kulsrud,wessonTok,freidberg2014}) (iii) $\mathbf{j}\times\mathbf{B}=\nabla p$, where $\mu_0\mathbf{j}=\nabla\times\mathbf{B}$ is the current density. It is possible to generalise the treatment beyond MHD with isotropic pressure \citep{rodriguez2021anis}, but we do not do that here.
\par
We understand an equilibrium solution to be the field $\mathbf{B}$ (and its associated flux surfaces $\psi$) as a function of $\mathbf{r}\in\mathbb{R}^3$, the position in the \textit{lab frame}, that satisfies this set of equations. This makes it natural to directly express $\mathbf{B}=\mathbf{B}(\mathbf{r})$ and $\psi=\psi(\mathbf{r})$, and solve our set of equations directly in this form. This approach is known as the \textit{direct-coordinate approach}, which was pioneered by \cite{mercier1962} and \cite{Solovev1970}, and subsequently developed and used by various authors \citep{shafranov1968influence,lortz_nuhrenberg_1978ballooning,Jorge_sengupta_landreman_2020_NF,jorge2020near,Duignan_Meiss2021normal_form,sengupta2024stellarator}. This approach is however not ideally suited to dealing with optimised stellarators. The reason is that many physical properties of the field are not a direct consequence of the field geometry, but rather $|\mathbf{B}|$ (see for example the neoclassical behaviour or single-particle dynamics \citep{boozer1983transport}), and thus an approach that incorporates $|\mathbf{B}|$ more organically is preferred. This brings us to the alternative option, which we favour in this paper, known as the \textit{inverse-coordinate approach} \citep{garrenboozer1991a,landreman2019}, in which $\mathbf{B}$ is described as a function of Boozer coordinates $\{\psi,\theta,\varphi\}$ \citep{d2012flux, boozer1981plasma}, where $\theta$ and $\varphi$ are the poloidal and toroidal angles respectively. 
\par
The use of Boozer coordinates is particularly convenient as it enables a succinct representation of $\mathbf{B}$ in a form that ensures equations (i) and (ii), 
\begin{subequations}
    \begin{align}
        \mathbf{B}=&\nabla\psi\times\nabla\theta+\iota(\psi)\nabla\varphi\times\nabla\psi \label{eqn:B_contra}\\
       =&G(\psi)\nabla\varphi+I(\psi)\nabla\theta+B_\psi(\psi,\theta,\varphi)\nabla\psi. \label{eqn:B_cov}
    \end{align} \label{eqn:def_B_field}
\end{subequations}
The first contravariant form introduces into the problem the rotational transform, $\iota$, while the latter brings the Boozer currents $I$ and $G$, as well as the covariant $B_\psi$. The solution of our problem is now a consistent set of functions $\{\iota,G,I,B_\psi\}$ as functions of Boozer coordinates.
\par
However, we are overlooking a key element in the construction. By adopting a representation in Boozer coordinates, the inverse-coordinate approach loses a direct link to real space, the lab frame. If one was given the field at some coordinate triplet $\{\psi_1,\theta_1,\varphi_1\}$, one would actually not know where in space this point is. The solution is thus incomplete. To connect Boozer coordinates to real space we define the following auxiliary functions $\{X,Y,Z\}$, such that
\begin{equation}
    \mathbf{r}=\mathbf{r}_\mathrm{axis}+X\hat{\pmb \kappa}+Y\hat{\pmb \tau}+Z\hat{\pmb t}, \label{eqn:inverse-coordiantes_FS}
\end{equation}
where $\mathbf{r}_\mathrm{axis}$ describes the magnetic axis of the equilibrium, and the triad $\{\hat{\pmb t},\hat{\pmb \kappa}, \hat{\pmb \tau}\}$ is the Frenet-Serret triad of the axis \citep{frenet1852courbes, animov2001differential}. That is, the unit tangent, normal and binormal vectors respectively. The complete description of the equilibrium requires solving for these space functions $\{X,Y,Z\}$ alongside the field; it will then be through Eq.~(\ref{eqn:inverse-coordiantes_FS}) that the shape of flux surfaces in space will be described (think of $\mathbf{r}$ at constant $\psi$). The functions $\{X,Y,Z\}$ are also directly involved in Eqs.~(\ref{eqn:def_B_field}). This comes apparent when using the dual relations $\partial_{q_i}\mathbf{r}=\mathcal{J}\epsilon_{ijk}\nabla q_j\times\nabla q_k$, for $\pmb q$ Boozer coordinates \citep{garrenboozer1991a,hazeltine2003plasma,d2012flux}. 
\par
With the problem set-up in this form, we may now turn to the asymptotic treatment. In the context of the near axis, the ordering `parameter' of the asymptotic treatment is some measure of the distance from the magnetic axis, which we will have to construct and shall refer to as $r$. The near-axis description consists on an expansion of the problem (i.e., the functions and governing equations) in powers of $r$ and the ordered solution of the ensuing hierarchy. Given that this distance $r$ is more a coordinate than an externally imposed constant, one may always choose a distance sufficiently close to the axis in which the asymptotic description holds. This distinguishes the near-axis from other similar asymptotic approaches such as large aspect-ratio expansions \citep{freidberg2014,greene_Johnson_welmer_1971tokamak,Solovev1970,cowley_kaw_1991analytic_high_beta}. 
With the closed axis enclosing zero toroidal flux $\psi=0$, it is natural to choose as a pseudo-radial coordinate $r=\sqrt{2\psi/\bar{B}}$, where $\bar{B}$ is a normalising reference magnetic field (assuming $\psi\geq0$). 

\subsection{Shaping the magnetic axis} \label{sec:nae_equilibrium_axis}
Before explicitly looking at the expansion in $r$, we must start with the choice of an appropriate reference axis. Its shape is rather important, as it does not only serve as reference, but also provides the basis in Eq.~(\ref{eqn:inverse-coordiantes_FS}) for the construction of the field. Its Frenet-Serret (FS) framing is chosen as a convenient basis \citep{serret1851quelques,frenet1852courbes,mathews1964mathematical,animov2001differential}, but we must comment on the subtleties that arise with that choice. Three-dimensional non-intersecting (\textit{simple}) space curves with nowhere vanishing tangent (\textit{regular}) \citep{moffatt1992helicity,oberti2016torus} nor curvature (\textit{geometric} \citep[Def.~4.2.4]{fuller1999geometric} or \textit{non-degenerate} \citep{pohl1968self} \textit{curves}), have a uniquely defined FS frame everywhere along the curve satisfying \citep{mathews1964mathematical},
\begin{equation}
    \frac{\mathrm{d}\mathbf{r}_\mathrm{axis}}{\mathrm{d}\ell}=\hat{\pmb t},\quad
    \frac{\mathrm{d}\hat{\pmb t}}{\mathrm{d}\ell}=\kappa\hat{\pmb \kappa},\quad
    \frac{\mathrm{d}\hat{\pmb \kappa}}{\mathrm{d}\ell}=-\kappa\hat{\pmb t} + \tau \hat{\pmb \tau},\quad
    \frac{\mathrm{d}\hat{\pmb \tau}}{\mathrm{d}\ell}=-\tau\hat{\pmb \kappa},  \label{eqn:FS_eqns}       
\end{equation}
where $\kappa$ and $\tau$ are the curvature and torsion of the curve, and $\ell$ is the length along it (with all $\hat{\cdot}$ vectors being normalised). All quantities to the right hand side are generally functions of $\ell$.
\par
Although this set of space curves exhausts the possibilities for QS fields \citep{landreman2019,rodriguez2022phases}, the construction of equilibria with poloidally closed $|\mathbf{B}|$ contours requires that curvature vanishes somewhere along the axis. The reason is that any amount of field line bending (in this case the magnetic axis) necessarily supports a finite gradient of $|\mathbf{B}|$ normal to the axis, $\nabla_\perp \ln B=\kappa\hat{\pmb\kappa}$, and thus some non-zero poloidal variation \citep[Eq.~(2.21.5)]{wessonTok}. 
We must then specialise on \textit{degenerate} curves with vanishing curvature points, which may be called \textit{flattening points} and shall take to be smooth. Although we shall assume such points to be isolated, they can be more or less flat, depending on how many $\ell$-derivatives of $\kappa$ vanish there. The order of the first non-vanishing derivative is called the \textit{order of the zero}. We shall additionally specialise on flattening points that are also points of stellarator symmetry \citep{dewar1998stellarator}, which simplifies the treatment significantly.
\par
In such curves, the FS frame is discontinuous at flattening points. If the order of the zero is even, though, one may show that the frame has a removable discontinuity there. If it is instead of odd order, the frame undergoes a $\pi$-rotation about the axis across the point. To form a continuous frame, then, a signed frame needs to be defined (the $\beta$-frame of \cite{carroll2013improving, plunk2019direct}). The frame is constructed by solving the regular FS equations, Eqs.~(\ref{eqn:FS_eqns}), starting from some non-degenerate point on the curve and moving along the curve, flipping the FS frame every time a zero of odd order is traversed. The result is a smooth, continuous frame in $\ell\in[0,L)$, where $L$ is the total length of the axis and the binormal and normal (as well as the curvature) are $\pm$ their FS form. Note that this signed frame still satisfy the FS equations (so Eq.~(\ref{eqn:inverse-coordiantes_FS}) and (\ref{eqn:FS_eqns}) can be interpreted to involvethe signed frame), and thus the near-axis construction in its usual form, as in LS, may be used. We will consider  the signed frame throughout this paper. See Appendix~\ref{app:ss_space curves} for details and proofs of these statements.
\par
Although continuity and differentiability of the frame is guaranteed as we move along the axis, because the curve is closed, this signed frame is not guaranteed to be periodic. In fact, if the sum of the order of all the flattening points is odd, then the frame at 0 and $L$ will have a flip respect to each other. In that case, we refer to the axis as a \textit{half helicity} curve, reminding ourselves the definition of the helicity as "the number of times the normal to the axis encircles the axis in a complete toroidal turn" (see Appendix~\ref{app:ss_space curves} for more precise definitions and \cite{Camacho2023helicity}). In practice, for a field that has a single magnetic well per field period and is QI, the curve will have two distinct flattening points: one where $B_0=B_\mathrm{min}$ and another where $B_0=B_\mathrm{max}$ (see Section~\ref{sec:2nd_order_qi_conditions} for the specific implications of QI that lead to this). The flattening point at $B_\mathrm{min}$ must be odd by omnigeneity, which makes the order of the top decide the \textit{helicity} of the axis: \textit{half helicity} if even, \textit{integer helicity} if odd. Because the frame itself is non-periodic in the half helicity case, the magnetic field described in Eq.~(\ref{eqn:inverse-coordiantes_FS}) would be unphysical unless the functions $X,~Y$ and $Z$ are made half-periodic as well. That is, functions that are periodic in the extended $\ell\in[0,2L)$ domain. A consistent treatment of these half-periodic axes is possible and explicitly discussed in detail in Appendix~\ref{app:cont_half_period}, where the necessary subtle modifications to the second order construction are detailed.
\par
The choice of a space curve with appropriate stellarator symmetric flattening points as our magnetic axis is the starting point of the near-axis construction. Some of these curves may be constructed straightforwardly following the prescriptions in \cite{plunk2019direct,rodriguez2022phases,jorge2022c,camacho-mata-2022,Camacho2023helicity}. A direct way of doing so by specifying the curvature and the torsion will be presented in a future publication. 
\par
Once we have a magnetic axis, we must specify how the magnetic field magnitude varies along it; namely, we must provide a $L$-periodic function $B_0(\ell)$. The same way as the field strength in a straight magnetic mirror can be externally curated, the function $B_0(\ell)$ must also be provided to the construction. We shall specialise, for simplicity, on fields with a single distinct trapping well along the axis, modulo the number of field periods, $N$, which will therefore have a single $B_\mathrm{min}$ and $B_\mathrm{max}$ (with the ratio $\Delta=(B_\mathrm{max}-B_\mathrm{min})/(B_\mathrm{max}+B_\mathrm{min})$ defined as \textit{mirror ratio}). As mentioned above, these extremal points must match the flattening points of the axis. This is a necessary consequence of having pseudosymmetric poloidal contours  \citep{rodriguez2023higher,skovoroda2005}. Due to stellarator symmetry, $B_0$ must be an even function about these points. 
\par
In order to sustain such a magnetic field line, we must thread the magnetic axis, through Ampere's law, with a poloidal current $G_0$ \citep[Eq.~(6.6.2)]{d2012flux}. Finally, and because the axis itself is a magnetic field line, there is a strict connection between the length along the field line $\ell$ and $\varphi$ (the Boozer angle), by virtue of being straight field line coordinates, $\mathrm{d}\ell/\mathrm{d}\varphi=|G_0|/B_0$ (see Eq.~(A20) in LS).  

\subsection{First order near-axis construction}
\begin{figure}
    \centering
    \includegraphics[width=0.9\textwidth]{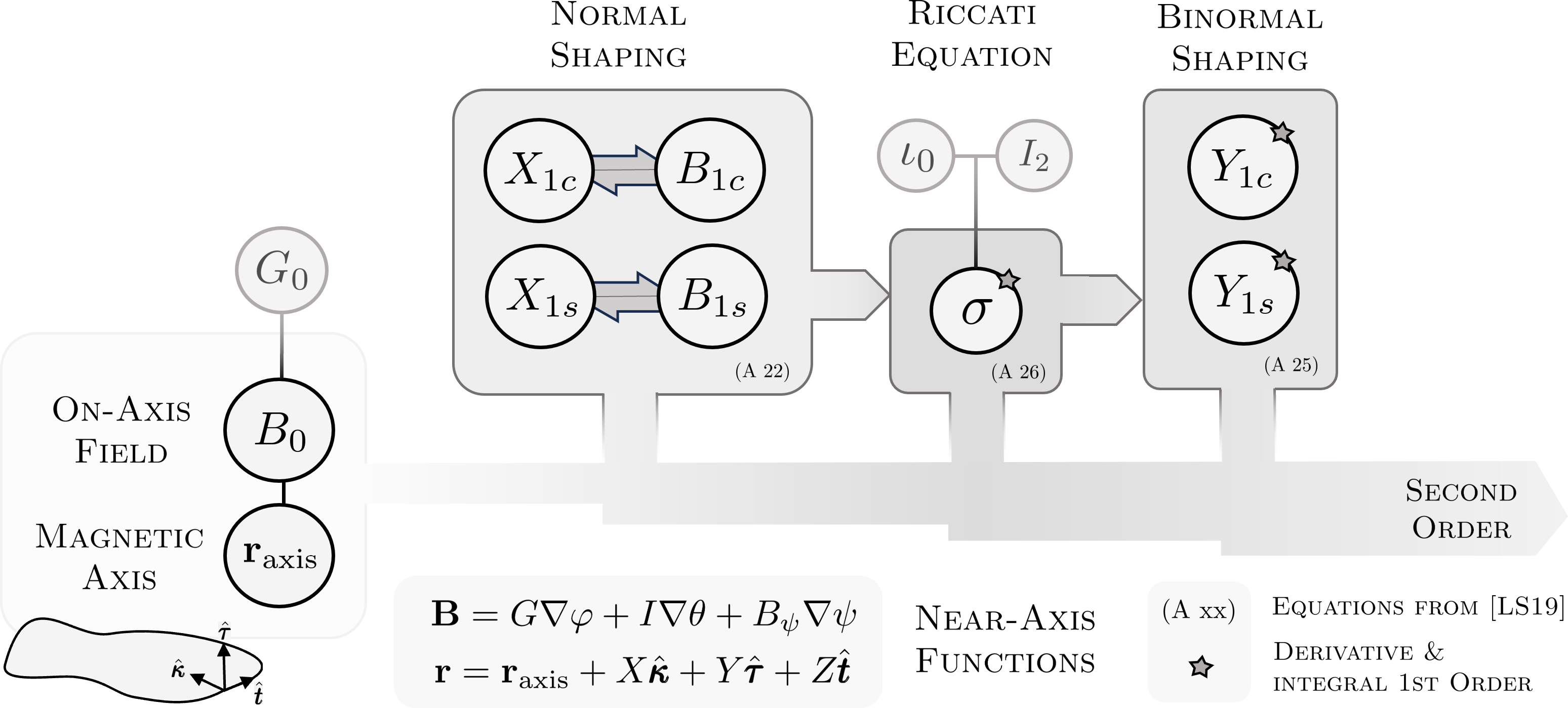}
    \caption{\textbf{Inverse-coordinate near-axis construction to first order.} Diagram depicting the key elements in the near-axis description of an equilibrium to first order, and the sequential order (left to right) in which to proceed. The construction starts with a shape for the magnetic axis and the magnetic field strength along it, both of which are given as inputs. To construct the neighbouring flux surfaces one must then provide the leading order variation of the normal shaping $X_1$, which directly relates to the field $B_1$ (this can go both ways). With that, one may then solve a differential equation on an auxiliary function $\sigma$, which is all that is needed to complete the flux surface description through $Y_1$. The encircled functions are the functions and parameters involved in the near-axis description, with the label (A xx) denoting the equations from \cite{landreman2019} needed to find them.}
    \label{fig:1st_order_nae_equilibrium}
\end{figure}

Once we have our axis, we move on to a description of the field in its neighbourhood; that is, we must now explicitly consider the expansion in $r=\sqrt{2\psi/\bar{B}}$, and look at the leading $O(r)$ parts of the field. Because of its radial-like nature, to avoid coordinate singularities on axis, the expansion in $r$ requires a careful coupling between powers of $r$ and harmonics of $\theta$ \citep{mercier1964equilibrium,kuo1987numerical,landreman2018a}. In particular, any function $f$ must take the form $f=\sum_{n=0}^\infty r^n f_n$ and $f_n=\sum_{m=0}^n (f_{nm}^c(\varphi)\cos m\chi+f_{nm}^s(\varphi)\sin m\chi)$, where the latter sum is over even or odd numbers depending on the parity of $n$, and $\chi=\theta-M\varphi$ with $M\in\mathbb{Z}$ the helicity of the signed frame. This defines the subscript notation to be repeatedly utilised throughout the paper\footnote{We may use the following shorthand as well: $F_{11}^c=F_{1c}$, $F_{11}^s=F_{1s}$, $F_{22}^c=F_{2c}$, $F_{22}^s=F_{2s}$ for any function that $F$ may be.}. Considering lower powers of $r$ then implies keeping a small number of poloidal harmonics, which is the key to the strength of the method. Resolving the resulting equations in $\chi$-harmonics and powers of $r$, the problem reduces to a hierarchy of equations on $\varphi$. The detailed accounts of the formal expansion of Eqs.~(\ref{eqn:def_B_field}) and equilibrium can be found in many works, in particular LS (originally in \cite{garrenboozer1991a}). We now present the structure of the problem to first order in $r$, including the equations that need to be solved, their physical meaning and the inputs needed to complete the description. The solution of the problem to $O(r)$ is summarised in a schematic way in Figure~\ref{fig:1st_order_nae_equilibrium}, which we shall follow closely in the description that follows. 
\par
We already started (leftmost part of Figure~\ref{fig:1st_order_nae_equilibrium}) the near-axis construction by the choice of an appropriate magnetic axis shape and the magnetic field strength on it. Naturally, when moving to $O(r)$ the field gains a finite flux surface build around the axis. Because of the analyticity constraint on the functions $X$ and $Y$ of Eq.~(\ref{eqn:inverse-coordiantes_FS}), these surfaces must be elliptical in the FS frame. There is a certain amount of freedom available (although limited) in the problem to choose how to shape this elliptical surface (see Figure~\ref{fig:1st_order_nae_equilibrium}). In particular, we must provide the function $X_1$ which specifies the distance along the normal from the flux surface to the axis as a function of $\chi$ and $\varphi$. It is convenient to define it in terms of two $\phi$-periodic functions \citep{garrenboozer1991a,plunk2019direct} $\bar{d}(\varphi)>0$ and $\alpha_1(\varphi)$, such that\footnote{It is important to note that by requiring $\alpha_1$ to be periodic and defining $\chi$ in terms of the helicity of the axis, the angle $\theta$ gains its poloidal meaning. Unless we define it this way, $\theta$ (whose meaning in real space comes through Eq.~(\ref{eqn:inverse-coordiantes_FS})) would generally be helical (see some further discussion in Appendix~\ref{app:cont_half_period}). Although the near-axis construction would remain valid in that case, the interpretation of key equilibrium quantities such as the rotational transform, Eq.~(\ref{eqn:B_cov}, would change.}
\begin{equation}
    X_1=\bar{d}\cos\left(\chi-\alpha_1\right). \label{eqn:X1}
\end{equation}
\par
The shaping $X_1$ is of particular physical significance: bringing the surface closer (further) to (from) the axis will (with a fixed gradient of $B$, from $\nabla_\perp(B^2/2)=B^2\pmb\kappa$) lead to a larger (lower) magnetic field strength on the surface. Thus the shaping brings in a direct control of the on-surface variation of $|\mathbf{B}|$. Within the context of the near axis, this formally translates to $B_1=\kappa B_0X_1$ (Eq.~(A22) of LS). Note that this formulation is different from the standard one, where the magnetic field magnitude is provided as a control input to the problem \citep{garrenboozer1991b,landreman2019, rodriguez2023constructing, landreman2022mapping}. We need to do so because of the presence of `insensitive' straight sections in which $B_1$ has an \textit{intrinsic} form (in particular, $B_1=0$ wherever $\kappa=0$). A similar observation will appear at higher orders. 
\par
Specifying the distance from the surface to the axis along the normal only constitutes one part in the description of the shape of flux surfaces. To complete it, we must know the behaviour along the binormal as well. Finding that $Y_1$ constitutes the last steps of the first order near-axis construction (see Figure~\ref{fig:1st_order_nae_equilibrium}). To find it, we must remember that we are describing a flux surface, and thus any cross-section of the surface must be thread by a constant toroidal flux. This means that as the field strength changes in $\varphi$, so must the cross-sectional area, so that $X_1Y_1\sim 1/B_0$. This justifies the condition $\bar{d}>0$, as flux surfaces would become infinitely elongated in the event of $\bar{d}$ ever vanishing. The exact relation between $X_1$ and $Y_1$ is however not so simple, because besides the conservation of flux, one must make sure that the resulting surface is consistent with the solenoidal magnetic field that lives on it. The consequence of this careful balance is a strong relation between the shaping of the elliptic flux surfaces (including both the elongation and the rotation), the shape of the axis and the rotational transform, $\iota$. 
\par
At a formal level, this careful balancing act reduces to the solution of a first order non-linear Riccati \citep[Sec.~1.4]{polyanin2017handbook} differential equation (Eq.~(A26) in LS) on a  periodic auxiliary function $\sigma(\varphi)$ (taken in the stellarator symmetric case to satisfy $\sigma(0)=0$) and the rotational transform on axis, $\iota_0$,
\begin{equation}
    \frac{\mathrm{d}\sigma}{\mathrm{d}\varphi}+\left(\bar{\iota}_0-\frac{\mathrm{d}\Tilde{\alpha}}{\mathrm{d}\varphi}\right)\left[\left(\bar{d}^2\frac{ B_0}{\bar{B}}\right)^2+1+\sigma^2\right]-2\bar{d}^2\frac{ B_0}{\bar{B}}\frac{\mathrm{d}\ell}{\mathrm{d}\varphi}\left(\frac{I_2}{\bar{B}}-\tau\right)=0. \label{eqn:sigma_eqn}
\end{equation}
The rotational transform is a scalar parameter that must be chosen to guarantee periodicity of $\sigma$ depending on the toroidal current $I_2$ assumed (often set to zero) \citep{landreman2018a}.\footnote{In the context of the direct-coordinate near-axis approach, no such non-linear equation is obtained, and one does instead get Mercier's expression for the rotational transform on axis \citep{mercier1964equilibrium}\citep[Eq.~(44)]{helander2014theory}. This is one of the advantages of the direct approach, which brings a tighter control and understanding of the geometry of the field. In the inverse-coordinate approach, it is customary to solve for $\iota_0$ and set $I_2$ to the desired value. However, the role of these two scalars may be reversed.} This equation has been explored in detail by other authors \citep{landreman2019direct, rodriguez2023constructing} and $\sigma$ given a geometric interpretation of (roughly) signifying the rotation of the ellipses respect to the FS frame \citep{rodriguez2023mhd,camacho-mata-2022}. Once the $\sigma$-equation is solved, $Y_1$ can be directly found (Eq.~(A25) in LS),
\begin{equation}
    Y_1=\frac{\bar{B}}{\bar{d}B_0}\left[\sin(\chi-\alpha_1)+\sigma\cos(\chi-\alpha_1)\right]
\end{equation}
and the first order construction is completed.

\subsection{Second order near-axis construction} \label{sec:nae_equilibrium_2nd}
\begin{figure}
    \centering
    \includegraphics[width=\textwidth]{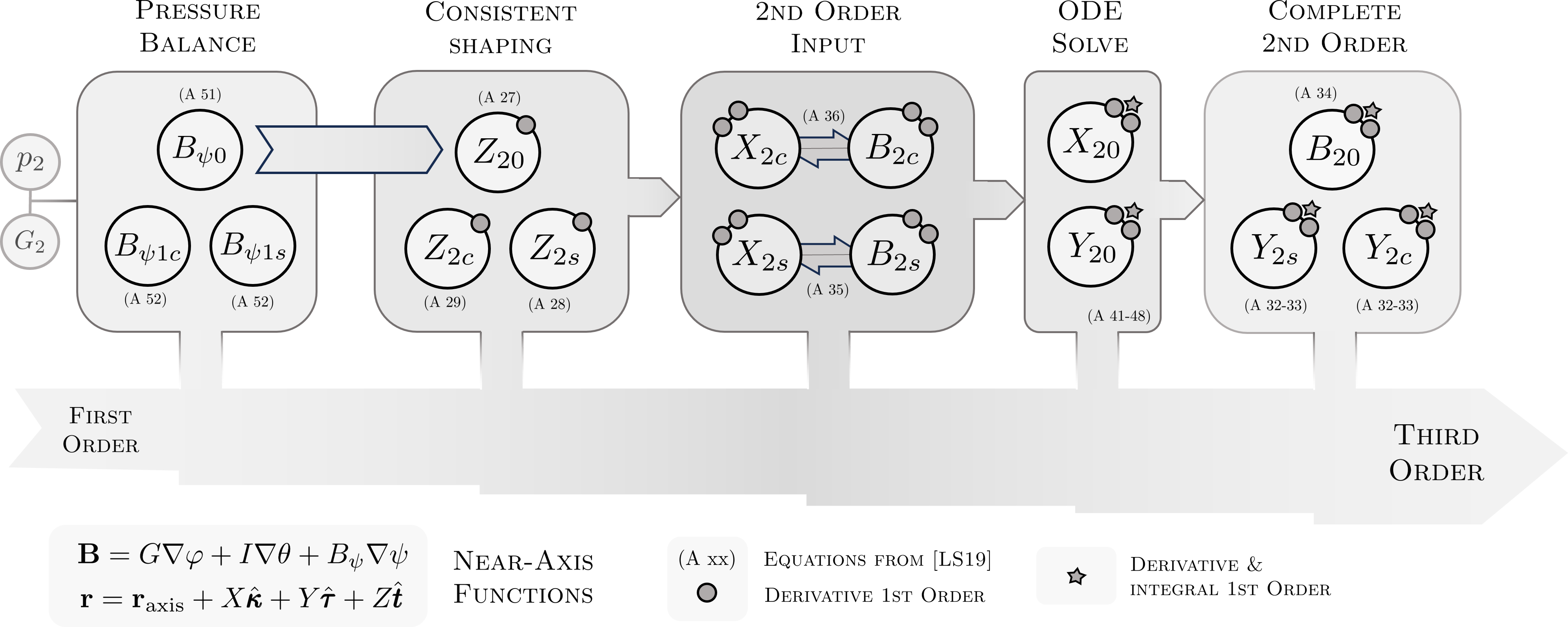}
    \caption{\textbf{Inverse-coordinate near-axis construction at second order.} Diagram representing the key elements in the typical second order near-axis equilibrium construction (left to right) to be taken as continuation of Fig.~\ref{fig:1st_order_nae_equilibrium}. The construction at second order starts by introducing the pressure gradient, $p_2$, explicitly, which $B_\psi$ and the current $G_2$ must balance. The $Z_2$ shaping is then uniquely determined to remain consistent. Providing the $\theta$-dependent shaping of $X_2$ as input (or the, for poloidal $|\mathbf{B}|$ fields less convenient, alternative $B_2$), the rest of the construction is uniquely determined, much like at first order. First one solves two coupled ODEs for the rigid displacement of flux surfaces (i.e., for $X_{20}$ and $Y_{20}$), to finally complete the magnetic field and flux surface shaping in the binormal direction. The encircled functions are the key elements needed to describe the field, with the label (A xx) denoting the equations from \cite{landreman2019} needed to find these quantities. The small circles on the edge of the functions denote that a derivative of lower order quantities is required to compute the function. The star, in turn, denotes that the function is a solution of an ODE with a derivative of lower order quantities in its inhomogeneous term.}
    \label{fig:2nd_order_nae_equilibrium}
\end{figure}

Much of the second order construction can be understood as a natural continuation to the first order, where similar physics govern the way to proceed. There is however, as indicated in Figure~\ref{fig:2nd_order_nae_equilibrium}, one significant difference: for the first time in the construction, the description of the field involves the pressure gradient, $p_2$, directly. Sufficiently close to the axis, the magnetic field is force-free, which is why $p$ was not involved at first order. 
The fulfilment of pressure balance is guaranteed by accommodating the magnetic field component $B_\psi=\mathbf{B}\cdot\mathbf{e}_\psi$ \citep{boozer1981plasma}. Formally, this involves the solution of simple first order ODEs (Eqs.~(A51)-(A52) in LS), which may be done straightforwardly using periodic, vanishing endpoint boundary conditions. In addition, to balance a finite pressure gradient it is necessary to have a finite net current gradient. A formal statement of that is that $G_2+\iota_0 I_2$ (related to the radial variation of the net poloidal and toroidal currents), can be found algebraically at this point in the construction (Eq.~(A50) in LS).  Thus there is one free constant in the second order construction, with either $G_2$ being determined by $p_2$ or vice versa.  
\par
Once pressure is in the problem, we solve for $Z_2$ (see Figure~\ref{fig:2nd_order_nae_equilibrium}). Note that unlike $X$ and $Y$, this component of the flux surface shaping does not directly affect the projection of the cross sections in the plane normal to the axis. Instead, it describes the non-planar shape of cross-sections of constant $\varphi$. Although on axis the toroidal angle is directly related to the length of the axis, for $\varphi$ to remain a Boozer angle in the neighbouring flux surfaces, it must be appropriately modified (this is of course related to the appropriate shaping of field lines which are tied to Boozer coordinates). This is what $Z_2$ does. Formally, finding $Z_2$ only involves operations of near-axis quantities already known, including their toroidal derivatives (see Eqs.~(A27-29) in LS). 
\par
With the toroidal coordinate resolved, we now meet a step analogous to that with $X_1$ shaping at first order (see Figure~\ref{fig:2nd_order_nae_equilibrium}). We have the freedom to impose some shaping along the normal to the axis at second order. In particular, the second harmonic components $X_{2s}$ and $X_{2c}$, which affect directly the triangularity and up-down symmetry of the cross sections \citep{thesis,rodriguez2023mhd}. As it occurred at first order, the choice of this distance will, through the curvature, directly affect the corresponding harmonics of the second order magnetic field strength $B_2$ (Eqs.~(A35-36) in LS),
\begin{subequations}
    \begin{equation}
        B_{2c} = \kappa B_0X_{2c}-\hat{\mathcal{T}}_c+\frac{(B_{1c}^\mathrm{QI})^2}{2B_0}\cos2\alpha_1, \label{eqn:X2c}
    \end{equation}
    \begin{equation}
        B_{2s}=\kappa B_0X_{2s}-\hat{\mathcal{T}}_s+\frac{(B_{1c}^\mathrm{QI})^2}{2B_0}\sin2\alpha_1, \label{eqn:X2s}
    \end{equation} \label{eqn:geoNAEeqs}
\end{subequations}
\noindent where $\hat{\mathcal{T}}_c$ and $\hat{\mathcal{T}}_s$ are known functions of first order quantities (see Appendix~\ref{app:2nd_order_eqns}), and $B_{1c}^\mathrm{QI}=B_0 \bar{d}\kappa$. As it occurred at first order, at the straight sections of the field, the second order shaping has no effect on the behaviour of the field (this favours, once again, the choice of $X_2$ harmonics ($X_{2c}$ and $X_{2s}$) as inputs, as opposed to $B_2$ ones). At those points, the field becomes solely determined by the first order solution. Elsewhere, the first order solution also has a direct effect (physically from the need to preserve divergencelessness and tangent fieldlines), but present a larger degree of freedom. The choice $X_{2c}=0=X_{2s}$, which we will refer to as \textit{minimal shaping}, makes that underlying structure manifest at second order. 
\par
From general MHD equilibria, we know that physical effects such as the Shafranov shift should come about here, influenced by the pressure gradient but also, and to a large extent, by the self-consistent shaping of the field \citep[Sec.~3.7]{wessonTok}. This naturally brings us to the next step in the second order near-axis construction (see Figure~\ref{fig:2nd_order_nae_equilibrium}), which is to self-consistently solve for $X_{20}$ and $Y_{20}$. These two elements are measures of Shafranov shift in the normal and binormal direction (i.e. the relative rigid displacement of flux surfaces with radius) \citep{landreman2021a,thesis,rodriguez2023mhd}. To find these functions consistently with the rest of pieces of the construction requires the solution of two linear coupled first order ordinary differential equations on $\varphi$. These may be solved numerically in a straightforward way (Eqs.~(A41)-(A48) in LS), the considerations for half-helicity fields requiring us to work with half periodic functions. With that solution and the second order shaping inputs, the construction of the surface at second order is completed algebraically finding $Y_{2s}$ and $Y_{2c}$ to preserve, as we learnt at first order, the constant flux assumption as well as the appropriate shaping of the field lines over them (Eqs.~(A32)-(A33) in LS). 
\par
Finally, we must compute the total magnetic field strength on the surface, of which only the second harmonics had been directly found after providing the normal shaping of the flux surfaces. The missing component is $B_{20}$, i.e. the average $\psi$-derivative of $|\mathbf{B}|$ on the surface. Of course, we could not compute this before knowing the complete shaping of the surfaces, and that is why its evaluation comes at last. Its construction is algebraic (Eq.~(A34) in LS), similar to Eqs.~(\ref{eqn:X2c})-(\ref{eqn:X2s}). Despite coming last in our solution method, the element $B_{20}$ plays a key role in physics, for instance MHD stability \citep{landreman2020magnetic, rodriguez2023mhd} and particle precession \citep{rodriguez2023precession,rodriguez2024maximum}.
\par
This concludes the essential elements of the near-axis equilibrium construction at second order, for a field with poloidally closed $|\mathbf{B}|$ contours. The above holds regardless of whether the axis is half or integer helicity, though the former case does have subtleties related to periodicity and continuity, as discussed in Appendix~\ref{app:cont_half_period}.

\section{Quasi-isodynamicity to second order} \label{sec:2nd_order_qi_conditions}
So far, the construction of the field near the axis has focused on the basic structure of an equilibrium stellarator symmetric magnetic field with nested flux surfaces and poloidally closed contours of $|\mathbf{B}|$. These properties are not sufficient to ensure that field is \textit{omnigeneous} \citep{hall1975,Cary1997} in the neighbourhood of the axis, {\em i.e.} the net radial drift of trapped particles will generally be non-vanishing. To complete the description of the field near the axis we must impose additional conditions, which in practice place severe constraints on the magnetic field magnitude $|\mathbf{B}|$.
\par
In the presence of stellarator symmetry, for the cancellation of the radial drift to occur, $|\mathbf{B}|$ must satisfy a number of symmetry conditions. This set of conditions on near-axis functions was originally derived to first order in \cite{plunk2019direct}, using the so-called \textit{Cary-Shasharina construction} \citep{Cary1997}. In a recent paper \citep{rodriguez2023higher}, the necessary conditions for quasi-isodynamicity were extended to second order in the near-axis expansion employing a more physical construction. Here we do not prove these conditions again, but simply present them, focusing on their consequences when considered alongside equilibrium.
\par
Let us start by recalling that in a stellarator symmetric configuration we must choose the magnetic field on axis to respect stellarator symmetry $B_0(\varphi)=B_0(-\varphi)$. That is, the magnetic well along the field line is symmetric, where $\varphi=0$ (in the domain $\varphi\in[-\pi/N,\pi/N)$) corresponds to the bottom of the well and we consider a single well per field period. In order for the radial drift ($\mathbf{v}_d\cdot\nabla\psi\sim\partial_\theta B_1$) on each side of this well to cancel each other exactly, we require \citep{plunk2019direct,rodriguez2023higher}
\begin{subequations}
    \begin{align}
        d(\varphi)=-d(-\varphi), \\
        \alpha_1=\bar{\iota}\varphi-\frac{\pi}{2},
    \end{align}
\end{subequations}
where $d=\kappa\bar{d}$. Because by stellarator symmetry the curvature of the axis must have well defined parity about the bottom of the well, $\bar{d}$ must be even. From the first order field construction it must also be non-zero, $\bar{d}>0$. The signed curvature is then an odd function of $\varphi$ at the bottom of the well, and thus it must have an odd zero of curvature there (as used in Section~\ref{sec:nae_equilibrium_axis}). In addition to parity, the zero of curvature should be chosen to satisfy an additional condition to avoid breaking the poloidal topology of $|\mathbf{B}|$ contours (i.e., bringing in \textit{puddles}) at finite $r$ and thus violating omnigeneity (and in particular pseudosymmetry \citep{mikhailov2002,skovoroda2005}) in the vicinity of field extrema. To achieve this, the order of the zero of curvature, call it $v$, must be such that $d$ does not grow too strongly near the bottom of the $B_0$ well.  This requires (see the discussion in \cite{rodriguez2023higher}) $2v\geq u$ for $B_0'\sim\varphi^{u-1}$ at the bottom of the well, and only equal when $(u,v)=(2,1)$. This very argument requires the critical points of $B_0$ to match flattening points of the axis.\footnote{The requirement of having an odd zero of curvature is not necessary at $B_{0,\mathrm{max}}$ (around which the QI symmetries do not really apply), where one must nevertheless satisfy the conditions related to \textit{pseudosymmetry}. In particular, matching of $B_0'=0$ and axis flattening points holds at all critical points.}
\par
The other ingredient of QI at first order, the particular form of $\alpha_1$, brings in an important issue. Quantities depending on $\alpha_1$, including $B_1$, are apparently not periodic (see Eq.~(\ref{eqn:X1})) if $\alpha_1$ is not. In particular, unless $\iota=M$, the helicity of the axis, there will be a lack of periodicity across the edges of the domain (i.e., the top of the well). This limitation was recognised early in the construction of QI fields, and requires sacrificing omnigeneity in a finite region near the edges of the domain. These regions where a finite controlled deviation from QI is enacted through the function $\alpha_1$ (which as we saw in the equilibrium construction must be periodic) are called \textit{buffer regions} \citep{plunk2019direct,camacho-mata-2022}. There are multiple ways in which these buffer regions may be constructed, with various degrees of sophistication and smoothness associated to them. Defining the odd parity function $\tilde{\alpha}=\alpha_1-\pi/2$ (to preserve stellarator symmetry), $\tilde{\alpha}$ must be constructed so that $\tilde{\alpha}\approx\bar{\iota}\varphi$ in the central region and vanishes at the endpoints. The differences between different buffers arise in the details of how this is done. For the purpose of this paper, we shall consider two options. One we call \textit{standard}, introduced in \cite{camacho-mata-2022}, where a polynomial in $\varphi$ is used to make $\tilde{\alpha}$ match $2k+1$ derivatives with the ideal QI function at the bottom of the well, where $k>0\in\mathbb{N}$ parametrises the family of buffers. Despite its simplicity, this choice has the potential issue presented by its non-smooth behaviour at the tops of the well, which, leads to discontinuous behaviour at second order in the construction. We present a smooth alternative, which we refer to as the \textit{smooth} buffer, that is $C^\infty$. In this case, $\tilde{\alpha}$ is written as a finite Fourier sine series, with its first $2k+1$ derivatives matching those of the ideal QI scenario at the bottom of the well. The details of this and other constructions are presented in Appendix~\ref{app:cont_buff}, alongside the issues of continuity and smoothness. Typical values of the parameter $k$ correspond to $k=5$ (roughly analogous to the $k=3$ choice for the standard construction, see Table~\ref{tab:compare_buffer}).
\par
Acknowledging and controlling the breaking of omnigeneity at first order through the use of buffers, we may construct physical fields that are approximately QI to first order. In the central region of the well the quality of omnigeneity is particularly high, and thus it is reasonable to ask about satisfying omnigeneity to next order there. {Partial omnigeneity may be interpreted as the condition of no net radial drift for a part of the trapped particle population.} For simplicity in the treatment, let us for now assume that the field is exactly QI at first order (see a more realistic treatment that takes the finite deviations into account in Appendix~\ref{app:2nd_order_deviation}), and define 
\begin{equation}
    B_2=B_{20}(\varphi)+B_{2c}(\varphi)\cos2\chi+B_{2s}(\varphi)\sin 2\chi. \label{eqn:B2_def}
\end{equation}
A second order QI field must then satisfy \citep[Eqs.~(32a-c)]{rodriguez2023higher},
\begin{align}
    B_{2c}=B_{2c}^\mathrm{QI}\cos2\bar{\alpha}-B_{2s}^\mathrm{QI}\sin2\bar{\alpha}, \label{eqn:B2cQIgeo}\\
    B_{2s}=B_{2s}^\mathrm{QI}\cos2\bar{\alpha}+B_{2c}^\mathrm{QI}\sin2\bar{\alpha}, \label{eqn:B2sQIgeo}    
\end{align}
where $\bar{\alpha}=\bar{\iota}\varphi-\pi/2$ is the ideal QI form of $\alpha_1$, and
\begin{subequations}
    \begin{gather}
        B_{20}(\varphi)=B_{20}(-\varphi), \label{eqn:B20_qi} \\
        B_{2s}^\mathrm{QI}(\varphi)=-B_{2s}^\mathrm{QI}(-\varphi), \label{eqn:B2s_qi} \\
        B_{2c}^\mathrm{QI}(\varphi)=\frac{1}{4}\left(\frac{B_0^2 d^2}{B_0'}\right)', \label{eqn:B2c_qi}
    \end{gather} \label{eqn:qiCondSS}
\end{subequations}

\noindent where primes denote derivatives in $\varphi$. The first two conditions, Eqs.~(\ref{eqn:B20_qi})-(\ref{eqn:B2s_qi}), are statements of stellarator symmetry, and may be easily satisfied if we guarantee that everything in the problem has this symmetry. A straightforward analysis of the the terms in Eqs.~(\ref{eqn:X2c})-(\ref{eqn:X2s}) shows that the terms coming from first order have the right symmetry (see Appendix~\ref{app:2nd_order_eqns}), and that the symmetry at second order is respected so long as one chooses $X_{2c}$ ($X_{2s}$) as odd (even) functions about the minimum of the well.
\par
The third condition, Eq.~(\ref{eqn:B2c_qi}), is a more interesting one, as in this case $B_{2c}^\mathrm{QI}$ becomes fully determined by lower order quantities. We now consider how to apply it to the relevant equilibrium equations,  Eqs.~(\ref{eqn:geoNAEeqs}). For this comparison between QI and equilibrium, it is convenient to rewrite Eqs.~(\ref{eqn:geoNAEeqs}) in terms of $B_{2c}^\mathrm{QI}$, {\em etc.}, by taking linear combinations as follows,

\begin{subequations}
    \begin{align}
        B_{2c}^\mathrm{QI}&=B_{2c,\mathrm{min}}^\mathrm{QI}+\kappa B_0\tilde{X}_{2c}, \label{eqn:B2c_qi_eq}\\
        B_{2s}^\mathrm{QI}&=B_{2s,\mathrm{min}}^\mathrm{QI}-\kappa B_0\tilde{X}_{2s}, \label{eqn:B2s_qi_eq}
    \end{align}
\end{subequations}
where,
\begin{subequations}
    \begin{align}
        B_{2c,\mathrm{min}}^\mathrm{QI}&=\frac{(B_{1c}^\mathrm{QI})^2}{2B_0}-\left(\hat{\mathcal{T}}_c\cos 2\alpha_1+\hat{\mathcal{T}}_s\sin 2\alpha_1\right), \\
        B_{2s,\mathrm{min}}^\mathrm{QI}&=\hat{\mathcal{T}}_c\sin 2\alpha_1-\hat{\mathcal{T}}_s\cos 2\alpha_1,
    \end{align}
\end{subequations}
and $\tilde{X}_{2c}=X_{2c}\cos{2\alpha_1}+{X}_{2s}\sin{2\alpha_1}$ and $\tilde{X}_{2s}={X}_{2c}\sin{2\alpha_1}-X_{2s}\cos{2\alpha_1}$.
\par
A straightforward attempt to impose omnigenity would be to simply substitute Eqn.~(\ref{eqn:B2c_qi}) into Eqn.~(\ref{eqn:B2c_qi_eq}), yielding an algebraic equation for the free shaping function $\tilde{X}_{2c}$.  As hinted previously, however, the zeros of curvature complicate this, and a number of constraints at those points must be satisfied by the first order solution, depending on the order of those zeros, to ensure that shaping function $\tilde{X}_{2c}$ is non-singular.

To interpret what is happening, the omnigenity constraint imposed by Eq.~(\ref{eqn:B2c_qi}) can be thought of as ``repairing'' defects in omnigeneity that are induced by the first order solution. This requires specially shaping the second order solution, but at locations where the axis straightens, such shaping is incapable of repairing the defects in omnigenity. One must do so by reaching back to first order, which boils down to satisfying additional point-wise constraints on the first order solution. Namely,
\begin{equation}
    \left(\frac{\mathrm{d}}{\mathrm{d}\varphi}\right)^k\left[B_{2c,\mathrm{min}}^\mathrm{QI}-\frac{1}{4}\left(\frac{B_0^2 d^2}{B_0'}\right)'\right]=0, \label{eqn:gen_mmin_cond}
\end{equation}
where $0\leq k<v$. In short, looking at second order provides additional insight on the choices made at first order.  A similar situation applies for $B_{20}$, the average radial derivative of $B$ (see the work in \cite{rodriguez2024maximum} for a discussion of it). 
\par
Let us try to understand the contents of this condition by considering the leading order ($k=0$) implications of the constraint at the point of minimum $B_0$. In the scenario of satisfying QI exactly at first order, we must satisfy at $\varphi=0$,
\begin{equation}
    \frac{(B_{1c}^\mathrm{QI})^2}{2B_0}+\hat{\mathcal{T}}_c=\frac{1}{4}\left(\frac{B_0^2 d^2}{B_0'}\right)' \label{eqn:QI_condition_plus_equil}
\end{equation}
A local expansion in $\varphi$ (assuming the right hand side to vanish)\footnote{The QI piece on the right hand side of Eq.~(\ref{eqn:QI_condition_plus_equil}) will indeed vanish in most scenarios. Only in what was called the \textit{special puddle} scenario in \cite{rodriguez2023higher} it will not; that is, when the axis has a flattening point of first order and $B_0$ a non-zero second derivative at the bottom of the well. All other combinations yield either non-omnigeneous fields (in the sense of being unable to confine deeply trapped particles) or have a vanishing right hand side.} yields,
\begin{equation}
    \underbrace{\frac{\bar{d}''}{\bar{d}}\left(1+\frac{1}{e^2}\right)}_\mathrm{Stretching}+\underbrace{\frac{B_0''}{B_0}}_{\mathrm{Breathing}}+\underbrace{\left(\frac{\tau G_0}{B_0}\right)^2\left(1+\frac{3}{e^2}\right)}_{\mathrm{Rotation}}-\underbrace{\frac{4}{e^2}\left(\frac{G_0}{B_0}\right)^2\frac{I_2}{\bar{B}}\tau}_\mathrm{Extra~twist}=0, \label{eqn:local_qi_cond}
\end{equation}
where $e=\bar{B}/\bar{d}^2B_0$ is the elongation of the elliptic cross-section along the binormal at the bottom of the well and primes denote derivatives in $\varphi$. This balance shows a physical competition between different elements that shape the resulting magnetic field lines. Stretching (variations in elongation) and breathing (variations in the cross-sectional area) of flux surfaces, alongside their rotation and twist shape field lines, and thus must collaborate to shape $B$ in a particular way so as to be omnigeneous. In the important stellarator-like scenario of minimal toroidal current, it follows from the above that it is necessary for $\bar{d}''/\bar{d}<0$, meaning that the cross-section will tend to become increasingly elongated (in the binormal direction) away from the bottom of the well. The larger the rotation and variation of $B$, the larger this stretching of flux surfaces needs to be. 
\par
An additional $v-1$ constraints (where $v$ is the order of the zero of curvature) similar to Eq.~(\ref{eqn:local_qi_cond}) can be obtained by taking additional derivatives of Eq.~(\ref{eqn:QI_condition_plus_equil}) and matching functions order by order, as shown in Eq.~(\ref{eqn:gen_mmin_cond}). Once all such local requirements are satisfied, and this `straight' mirror-like section of the field is left (sufficiently far from flattening points), one may obtain a well-behaved function $\tilde{X}_{2c}$ to enforce the second order omnigenity constraint. However, this leaves us with little control on the amount of shaping at second order, which generally proves to be excessive (see some details on this way of approaching the problem in Appendix~\ref{app:alt_2nd_omni}).  If the emphasis is placed on minimising this shaping, we then have what we call the \textit{minimal shaping} approach.
\par
The idea is to look for the subset of first order fields for which one is capable of satisfying the omnigeneity condition at higher order, i.e. QI setting $X_{2s}=0=X_{2c}$. In that case, the lower order near-axis construction is fully responsible for the extent to which the field is more or less omnigeneous at second order. In practice this means that we need to choose $B_0$, $\bar{d}$ and the axis shape (as well as $\alpha_1$) so as to satisfy Eq.~(\ref{eqn:QI_condition_plus_equil}) in $\varphi$ (now not just at $\varphi=0$). 
\par
Although we have verbalised the problem in a rather straightforward way, the problem is generally a hard one to solve (given its non-linearity). It is thus natural to treat this as an optimisation problem where 0th and 1st order inputs are varied to seek minimisation of $f_\mathrm{QI}=\left(B_{2c,\mathrm{min}}^\mathrm{QI}-B_{2c}^\mathrm{QI}\right)^2$, where the latter is the ideal omnigeneous form in Eq.~(\ref{eqn:B2c_qi}). Optimisation of near-axis fields is not unfamiliar \citep{landreman2022mapping,jorge2022c,rodriguez2023constructing,Camacho2023helicity}, and known codes and techniques may be used to that end. Recalling the presence of a buffer region extending from the tops of the well, it is logical to confine the optimization of $f_\mathrm{QI}$ only to values of $\varphi$ where the deviation of $\alpha_1$ from its ideal QI value is small. This qualitative choice of a region of interest can be made more quantitative by comparing the implications of breaking omnigeneity at first order to breaking it at second order. Some of the details are presented in Appendix~\ref{app:2nd_order_deviation}. The main takeaway is nevertheless that wherever $\alpha_\mathrm{buf}$ is small, then the QI $B_{2c}$ criterion will dominate the behaviour, and thus one should enforce the QI condition there. We present an example of an optimised field following this construction in the next section.

\section{Numerical benchmark of second-order constructions} \label{sec:numerical_bench}
Having described the near-axis construction in detail in the preceding sections, we now present a number of numerical examples of near-axis fields with poloidal contours of $|\mathbf{B}|$ through (and including) second order. We do so for a variety of different inputs and benchmark the resulting constructions against the well established global equilibrium code \texttt{VMEC} \citep{hirshman1983} code. In that regard, we mirror the procedure in \cite{landreman2019}, which carried similar benchmarking work in the context of quasi-symmetric fields. The agreement between the global solutions and the near-axis description in the limit of large aspect ratio should be taken as evidence of both the correct implementation of the near-axis code and the correctness of the near-axis construction. This numerical check is necessary to set a solid foundation to the second order treatment of QI fields, so that it may be leveraged in future applications and studies. 
\par
To perform this numerical benchmark, we first need to construct the near-axis fields. We use numerical considerations like those spelled out in \cite{landreman2019} and implemented in \texttt{pyQSC}. In the context of QI fields this was originally implemented in \texttt{pyQIC}, used in \cite{jorge2022c}, which we have significantly extended to appropriately deal with the second order\footnote{The code may be found in \texttt{https://github.com/SebereX/pyQIC.git}, the right version and branch to use to be found in the Zenodo repository, and we refer to it for further details about its implementation.}. In particular, a careful treatment of smoothness, buffers and the difficulties posed by half helicity axes have been incorporated (these are the subject of some of the theoretical discussions touched upon in the main text and detailed in Appendices~\ref{app:cont_half_period} and \ref{app:cont_buff}). 
\par
Once we are able to numerically construct a near-axis field, we then proceed to make a comparison against global equilibria. We do so by constructing flux surface boundaries at different finite radii using the near-axis \citep{landreman2019} (including the necessary third order fixes), and using them as inputs to global equilibria of different aspect ratios in \texttt{VMEC} \citep{hirshman1983}.\footnote{This benchmark requires the solution in \texttt{VMEC} to be found with a high degree of angular resolution, in part, as a result of the ill-suitability of the cylindrical coordinate to describe these configurations. Details on the specifics may be found in the Zenodo repository associated to this paper, to which we refer the interested reader. As orientative figures, the code was run with a large angular resolution ($\texttt{mpol}=8$ and $\texttt{ntor}=30$) and extremely low error tolerances ($\texttt{ftol}=1$e$-18$).} The field near the axis may then be compared between the near-axis and global fields. To that end, we compute the magnetic field magnitude in Boozer coordinates using \texttt{BOOZXFORM} \citep{sanchez2000ballooning}, and by fitting the radial profiles of the various harmonics, construct estimates of $B_{0,\texttt{VMEC}}$, $B_{1,\texttt{VMEC}}$ and $B_{2,\texttt{VMEC}}$. A comparison of the latter to $B_2$ from within the near-axis framework and its behaviour across different aspect ratios, $A$, can then be used as a measure of correctness of the solution. We define,
\begin{equation}
    \Delta B_{\mathrm{rms}.x}=\sqrt{\frac{1}{2\pi}\int_0^{2\pi}\left(B_{x}-B_{x,\texttt{VMEC}}\right)^2\mathrm{d}\varphi}, \label{eqn:delta_B}
\end{equation}
for $x=20,~2s,~2c$. 
\par
At extremely large aspect ratios ($A>100$), the equilibrium solver starts to struggle finding solutions with the desired levels of accuracy, which hinders the comparison at such high values. The trend with $A$ can nevertheless be used to judge the level of agreement between the asymptotic description and the full solution.

\subsection{Minimally shaped vacuum configuration} \label{sec:katia_n2_buffer}
Let us start our benchmark by looking at a simple case which we take from the recent publication of \cite{camacho-mata-2022}. We consider the $N=2$ field in that paper, which is approximately QI at first order with a standard buffer region with $k=2$. The axis is defined to have an integer helicity with first order zeroes of curvature, parametrised by
\begin{subequations}
    \begin{align}
        R\,[\text{m}] &= 1-\frac{1}{17}\cos2\phi, \\
        Z\,[\text{m}] &= \frac{0.8}{2.04}\sin2\phi+\frac{0.01}{2.04}\sin4\phi,
    \end{align}
\end{subequations}
where the form of $R$ guarantees the correct behaviour of flattening points \citep{rodriguez2022phases,camacho-mata-2022}. The magnetic field on axis is simply described by $B_0\,[\text{T}]=1+0.15\cos2\varphi$, and $\bar{d}=0.73$ is assumed constant. We extend this field into second order following the minimal shaping assumption; that is, $X_{2c}=0=X_{2s}$. For this choice, the second-order field is a mere extension of the first order, and investigating the properties of this second order completion may be used in the future as a tool to investigate the properties and suitability of the first order near-axis construction, all within the near axis framework. 
\par
\begin{figure}
    \centering
    \includegraphics[width=\textwidth]{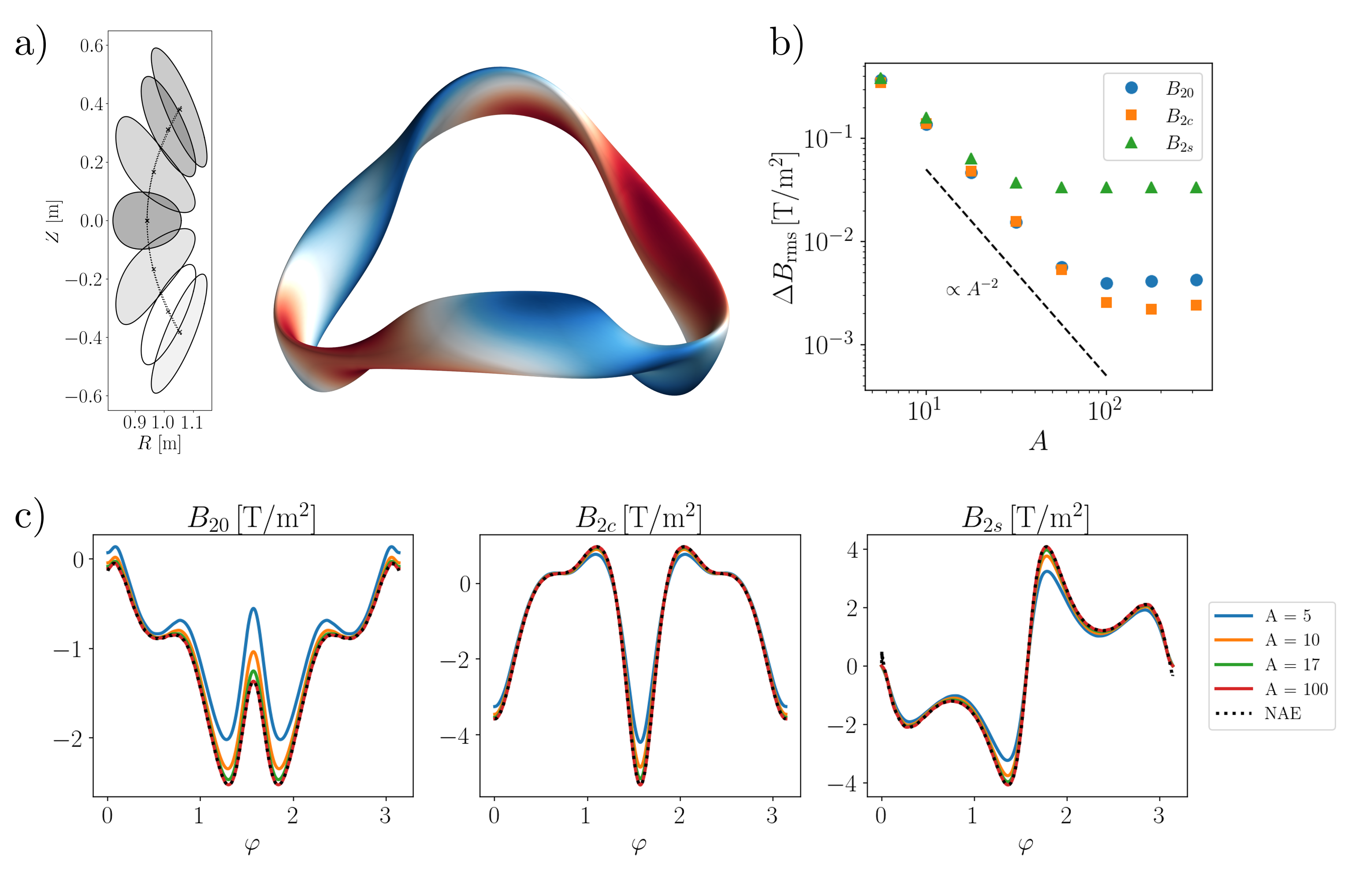}
    \caption{\textbf{Global equilibrium construction from the second-order near-axis field in Sec.~\ref{sec:katia_n2_buffer}.} The figure presents the second order near-axis construction based on the $N=2$ field of \cite{camacho-mata-2022} presented in Sec.~\ref{sec:katia_n2_buffer}. (a) Cross-sections at constant cylindrical angle in a half period and 3D rendering at $A=10$. The dotted line traces the magnetic axis, with crosses representing the intersection with the cross-sections. (b) Root-mean-square difference in the second order near-axis magnetic field between the ideal near-axis description and the finite aspect ratio global equilibrium construction with \texttt{VMEC}. A reference quadratic scaling $\propto 1/A^2$ is given, from which the convergence of the error deviated at larger aspect ratios. (c) Explicit comparison between the different poloidal components of the second order magnetic field magnitude from the global \texttt{VMEC} equilibrium at different aspect ratios, and the ideal near-axis value (black dotted line). }
    \label{fig:NFP2_Katia_std_buffer}
\end{figure}
The construction of the near-axis field and the comparison to the global equilibrium is shown in Figure~\ref{fig:NFP2_Katia_std_buffer}. The top left plots show cross-sections and a flux surface of the configuration evaluated at $r\,[\text{m}]=0.1$, which roughly corresponds to an aspect ratio of $A=1/r=10$. The bottom plot shows a direct comparison between the different components of $B_2$ as a function of $\varphi$ for a number of aspect ratios, showing that as the aspect ratio $A$ is increased, the agreement becomes better. Even at as low an aspect ratio as $A=5$ the second order near-axis construction reproduces the behaviour of the equilibrium at least qualitatively. This points towards the significance and physical nature of second order properties. The top right plot shows the level of agreement as a function of aspect ratio in a more quantitative fashion. It shows the root-mean-square difference between the \texttt{VMEC} and near-axis functions defined in Eq.~(\ref{eqn:delta_B}). Although there is an initial improvement with increasing aspect ratio at small $A$, this seems to saturate quite rapidly (at about $A\sim20$). The reason behind this discrepancy can be traced to the behaviour of $B_{2s}$ in Figure~\ref{fig:NFP2_Katia_std_buffer}c. Near the endpoints of the domain, one can see that the function $B_{2s}$ from the near-axis construction is actually discontinuous. And of course, the global equilibrium solve cannot reproduce such unphysical features. Hence the saturation of the error at a level dictated by the size of the discontinuity.
\par
\begin{figure}
    \centering
    \includegraphics[width=\textwidth]{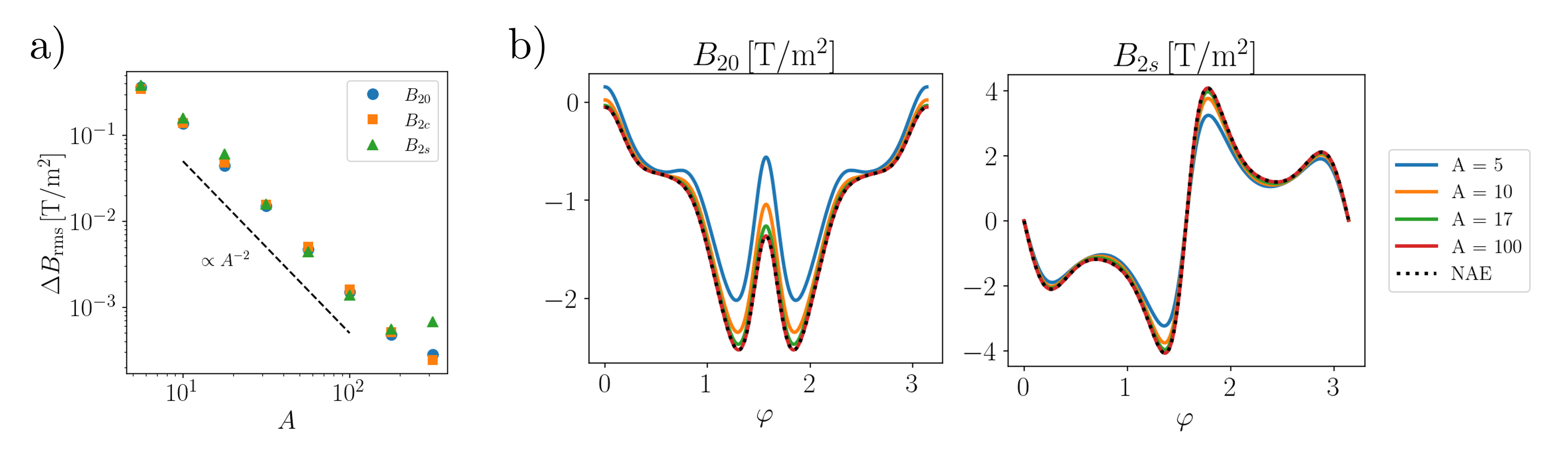}
    \caption{\textbf{Comparison of second order field of the equilibrium in Figure~\ref{fig:NFP2_Katia_std_buffer}  with smooth buffer regions.} The figure presents the comparison of the second order near-axis field for the same configuration as in Figure~\ref{fig:NFP2_Katia_std_buffer} but for the difference in the analyticity of the construction near the field edges. (a) Root-mean-square difference in the second order near-axis magnetic field between the ideal near-axis description and the finite aspect ratio global equilibrium construction with \texttt{VMEC}. A reference quadratic scaling $\propto 1/A^2$ is given, from which the convergence of the error deviated at larger aspect ratios. (c) Explicit comparison between the different poloidal components of the second order magnetic field magnitude from the global \texttt{VMEC} equilibrium at different aspect ratios, and the ideal near-axis value (black dotted line). The agreement near the edge is better than that of Figure~\ref{fig:NFP2_Katia_std_buffer} }
    \label{fig:NFP2_Katia_smooth_buffer}
\end{figure}

The origin of said discontinuity had already been discussed in the main text preceding this section: it is the lack of smoothness in the standard buffer. To show that this is responsible for the observed disagreement, we repeat the same numerical exercise, but this time using a smooth buffer with $k=5$. The results are shown in Figure~\ref{fig:NFP2_Katia_smooth_buffer}, where the previous saturation of the error disappears.
\par
The disagreement between the near-axis and global equilibrium scales like $\sim 1/A^2$ (this is the difference in the second order field $B_2$, and not $r^2B_2$). This quadratic reduction of the error is expected from a construction that is accurate to second order, as the finite build of the near axis is \citep{landreman2019}. This requires the error to be at least $1/A$, but because of analyticity of the field near the magnetic axis, the $m=0,2$ harmonics only exhibit even powers and thus the scaling with $1/A^2$. This behaviour is observed over a wide range of aspect ratios. At the highest values of aspect ratio (in Figure~\ref{fig:NFP2_Katia_smooth_buffer}, $A=316$) there appears to be some level of discrepancy as \texttt{VMEC} struggles to converge (note however the high demands we are placing on the code, as we require the field to be correct to about 1 part in $10^8$ or more). 
\par
With this first example we show that: (i) the near-axis construction to second order for QI fields works, (ii) that the non-smooth constructions cannot be faithfully reproduced by global physical equilibria, but (iii) that their qualitative aspects can.

\subsection{Shaped, finite plasma-$\beta$ configuration} \label{sec:beta_shape}

\begin{figure}
    \centering
    \includegraphics[width=\textwidth]{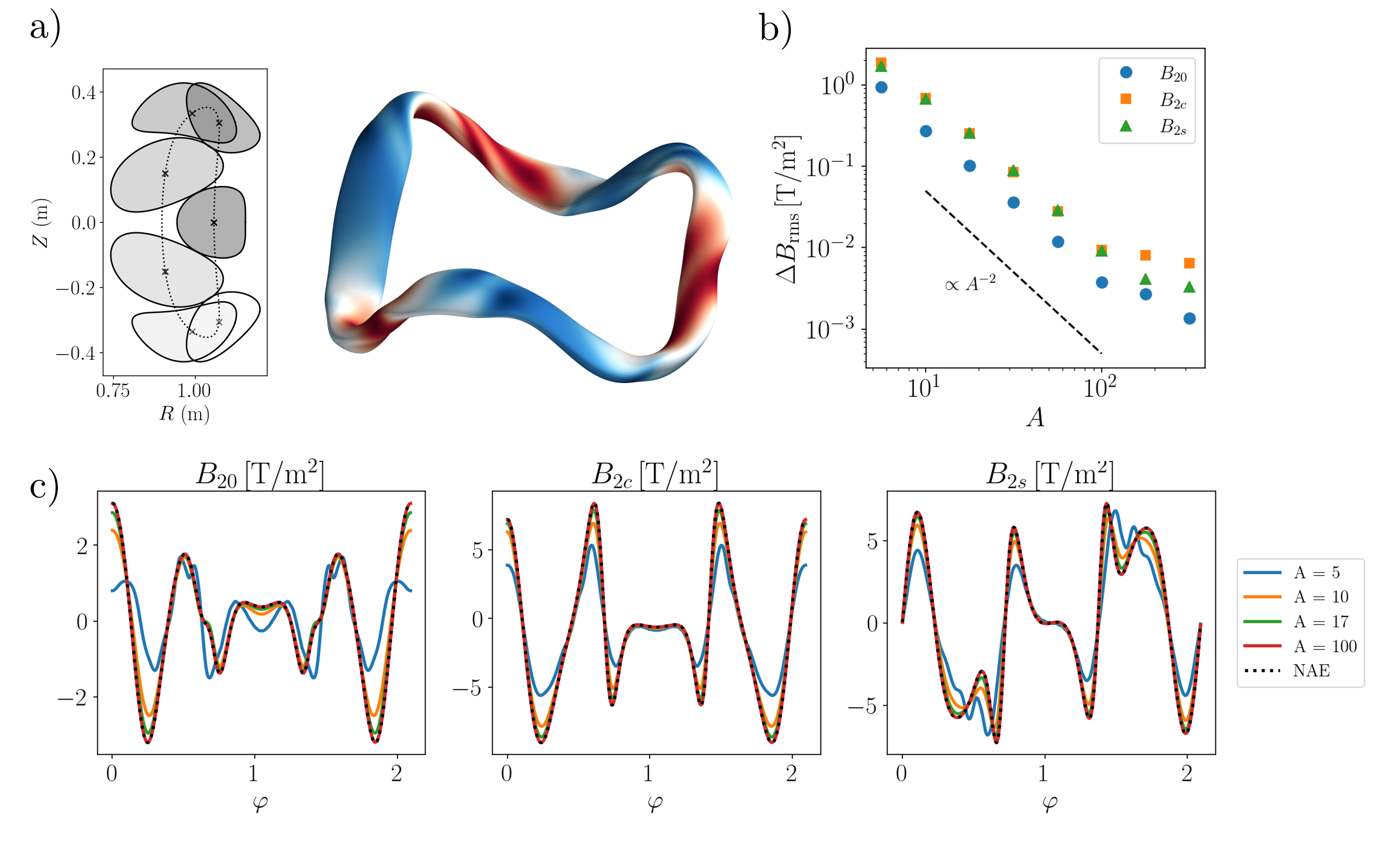}
    \caption{\textbf{Global equilibrium construction from the second-order near-axis field in Sec.~\ref{sec:beta_shape}.} The figure presents the second order near-axis construction based on \cite{camacho-mata-2022} presented in Sec.~\ref{sec:katia_n2_buffer} including finite $\beta$, toroidal current and additional second order shaping. (a) Cross-sections at constant cylindrical angle in a half period and 3D rendering at $A=10$. The dotted line traces the magnetic axis, with crosses representing the intersection with the cross-sections. (b) Root-mean-square difference in the second order near-axis magnetic field between the ideal near-axis description and the finite aspect ratio global equilibrium construction with \texttt{VMEC}. A reference quadratic scaling $\propto 1/A^2$ is given, from which the convergence of the error deviated at larger aspect ratios. (c) Explicit comparison between the different poloidal components of the second order magnetic field magnitude from the global \texttt{VMEC} equilibrium at different aspect ratios, and the ideal near-axis value (black dotted line). }
    \label{fig:NFP3_Katia_beta}
\end{figure}

We now consider a more involved example that includes finite plasma $\beta$ and non-zero eplicit second order shaping. In this case, we base the construction off the $N=3$ in \cite{camacho-mata-2022}, with a smooth buffer region. The axis is defined approximately by,
\begin{subequations}
    \begin{multline}
        R [\text{m}]= 1+ 0.09075485\cos 3\phi -0.02058279 \cos 6\phi -0.01106766\cos9\phi \\
        -0.001644390e \cos 12\phi
    \end{multline}
    \begin{equation}
        Z [\text{m}]= 0.36 \sin 3\phi + 0.02\sin 6\phi + 0.01\sin 9\phi,
    \end{equation}
\end{subequations}
which when implemented numerically one should carefully check to satisfy the vanishing curvature condition at the stellarator symmetric points. The magnetic strength on axis is $B_0\,[\text{T}]=1+0.25\cos3\varphi$, and $\bar{d}=0.73$ once again. In this case we choose some arbitrary non-zero values for $X_{2c}[\text{m}^{-1}]=0.1\sin3\varphi-0.6\sin6\varphi+0.1\sin9\varphi$ and $X_{2s}[\text{m}^{-1}]=0.1-0.1\cos6\varphi+0.6\cos9\varphi$. Finally, we put a finite toroidal current $I_2\,[\text{T/m}] = -0.9$ and pressure gradient $p_2\,[\text{Pa/m}^2]=-6\times10^5$. The resulting field and comparison to \texttt{VMEC} is presented in Figure~\ref{fig:NFP3_Katia_beta}.
Once again, the quadratic scaling of the error shows agreement between the near-axis construction and the global equilibrium solve. Note that this time the equilibrium is significantly more shaped than in the case of Section~\ref{sec:katia_n2_buffer}, in part as a consequence of the larger number of field periods (and thus larger toroidal derivatives) but also the explicit shaping introduced at second order. This increased complexity is apparent in the comparison of the different components of $B_2$, but also on the scaling of the error $\Delta B_\mathrm{rms}$ itself, as the deviations from the ideal scaling occur at lower aspect ratios. This case is an example of one of the struggles in the design of near-axis QI fields, which is to avoid this extreme shaping.

\subsection{Half helicity case} \label{sec:half_period_config}
The same systematic construction can be followed for the special QI class of fields with half-helicity axes. We construct an $N=3$ half helicity field with a smooth $k=5$ buffer. When dealing with half helicity axes, and especially when high control is required on the curvature and torsion, as is indeed needed in order to control the extreme shaping that is the tendency of QI fields, it is convenient to define the axis by directly parametrising the curvature and the torsion. In that case, one needs to reconstruct the axis by integrating Frenet-Serret (FS) equations, Eqs.~(\ref{eqn:FS_eqns}). The well-known drawback of doing so is that curvature and torsion must be chosen carefully to guarantee closure of the axis, which must be achieved numerically to a sufficiently high degree of accuracy. Failing to do so will lead to discontinuities (or lack of smoothness in the best of cases) in the near-axis construction of flux surfaces, and thus will lead to an unphysical field.\footnote{Note that this is an issue \textit{only} when it comes to constructing the field in real space. That is, the near-axis functions themselves within the near-axis framework do not see any lack of smoothness or discontinuity. It is only through Eq.~(\ref{eqn:inverse-coordiantes_FS}) that this manifests. In the comparison to \texttt{VMEC} we need to make sure that this is not a problem.} Taking this into account, we consider in this case, 
\begin{subequations}
    \begin{gather}
        \kappa [\text{m}^{-1}]=-\frac{8.0725268}{2}\left[1+\cos \left(6\pi\frac{\ell}{L}\right)\right]\sin\left(3\pi\frac{\ell}{L}\right)\sin\left(6\pi\frac{\ell}{L}\right), \\
        \tau [\text{m}^{-1}]= 1.34174997-0.133333\cos\left(6\pi\frac{\ell}{L}\right),
    \end{gather}
\end{subequations}
and $L [\text{m}]= 2\pi$, which gives a first order zero at the bottom of the well and second order at the top. The magnetic field $B_0$ and the shaping $\bar{d}$ are not presented explicitly in the text, but can be found in the repository associated to this paper, as they have been chosen carefully to limit the amount of elongation and shaping of the construction. The steps informing the choice of inputs and the use of the FS approach will be detailed in a future publication. We complete the inputs to the near-axis construction by choosing $X_{2c}=0=X_{2s}$.
\par
\begin{figure}
    \centering
    \includegraphics[width=\textwidth]{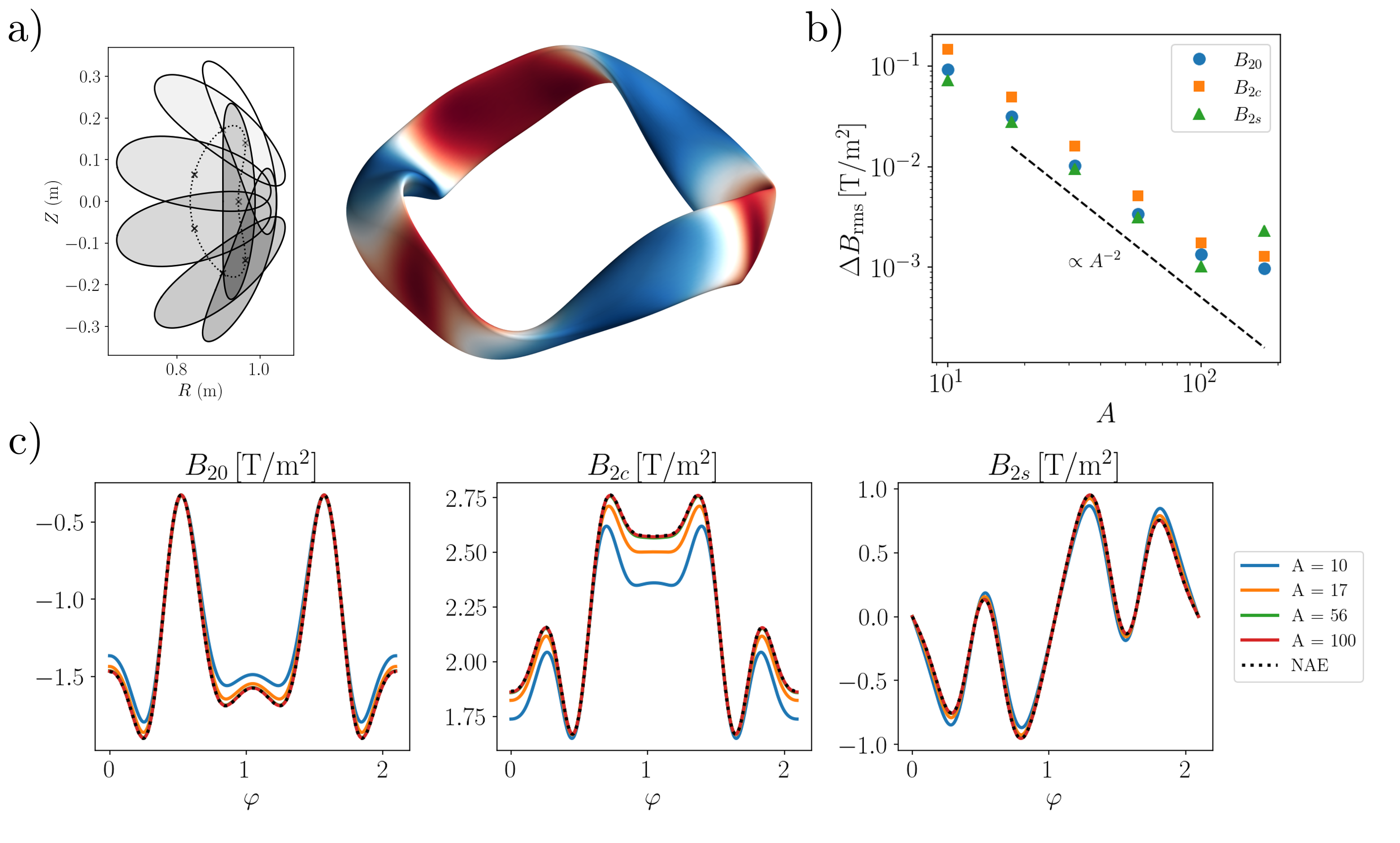}
    \caption{\textbf{Global equilibrium construction from the second-order near-axis field in Sec.~\ref{sec:half_period_config}.} The figure presents a second order near-axis construction for a half-helicity, three period field. (a) Cross-sections at constant cylindrical angle in a half period and 3D rendering at $A=10$. The dotted line traces the magnetic axis, with crosses representing the intersection with the cross-sections. (b) Root-mean-square difference in the second order near-axis magnetic field between the ideal near-axis description and the finite aspect ratio global equilibrium construction with \texttt{VMEC}. A reference quadratic scaling $\propto 1/A^2$ is given, from which the convergence of the error deviated at larger aspect ratios. The \texttt{VMEC} solver struggles at larger aspect ratios. (c) Explicit comparison between the different poloidal components of the second order magnetic field magnitude from the global \texttt{VMEC} equilibrium at different aspect ratios, and the ideal near-axis value (black dotted line). }
    \label{fig:NFP3_half_period}
\end{figure}
The results of the comparison to \texttt{VMEC} are shown in Figure~\ref{fig:NFP3_half_period}, where we get the expected agreement. This shows that the treatment of the half helicity scenario (as discussed in Section~\ref{sec:nae_equilibrium} and Appendix~\ref{app:cont_half_period}) and is indeed the correct one.

\subsection{QI optimised minimally shaped field} \label{sec:opt_QI_config}
Finally, we present a field which has been optimised to be omnigeneous at second order near the bottom of the well following Section~\ref{sec:2nd_order_qi_conditions}. In this proof-of-principle exercise, we allowed the $Z$ harmonics as well as $\bar{d}$ to vary as our degrees of freedom, and penalise deviations from omnigeneity at second order in the central 20\% of the domain. The starting point of the construction was the same as that of Section~\ref{sec:beta_shape}, namely the $N=3$ from \cite{camacho-mata-2022}. The resulting $Z$ component of the axis,
\begin{equation}
    Z [\text{m}] \approx 0.397890 \sin 3\phi + 0.0244030\sin 6\phi + 0.00292413\sin 9\phi,
\end{equation}
and 
\begin{multline}
    \bar{d} \approx 0.642233+0.000346819\cos 6\varphi+0.00114030\cos9\varphi+0.00366299\cos12\varphi\\
    -0.00230278\cos15\varphi.
\end{multline}

\begin{figure}
    \centering
    \includegraphics[width=\textwidth]{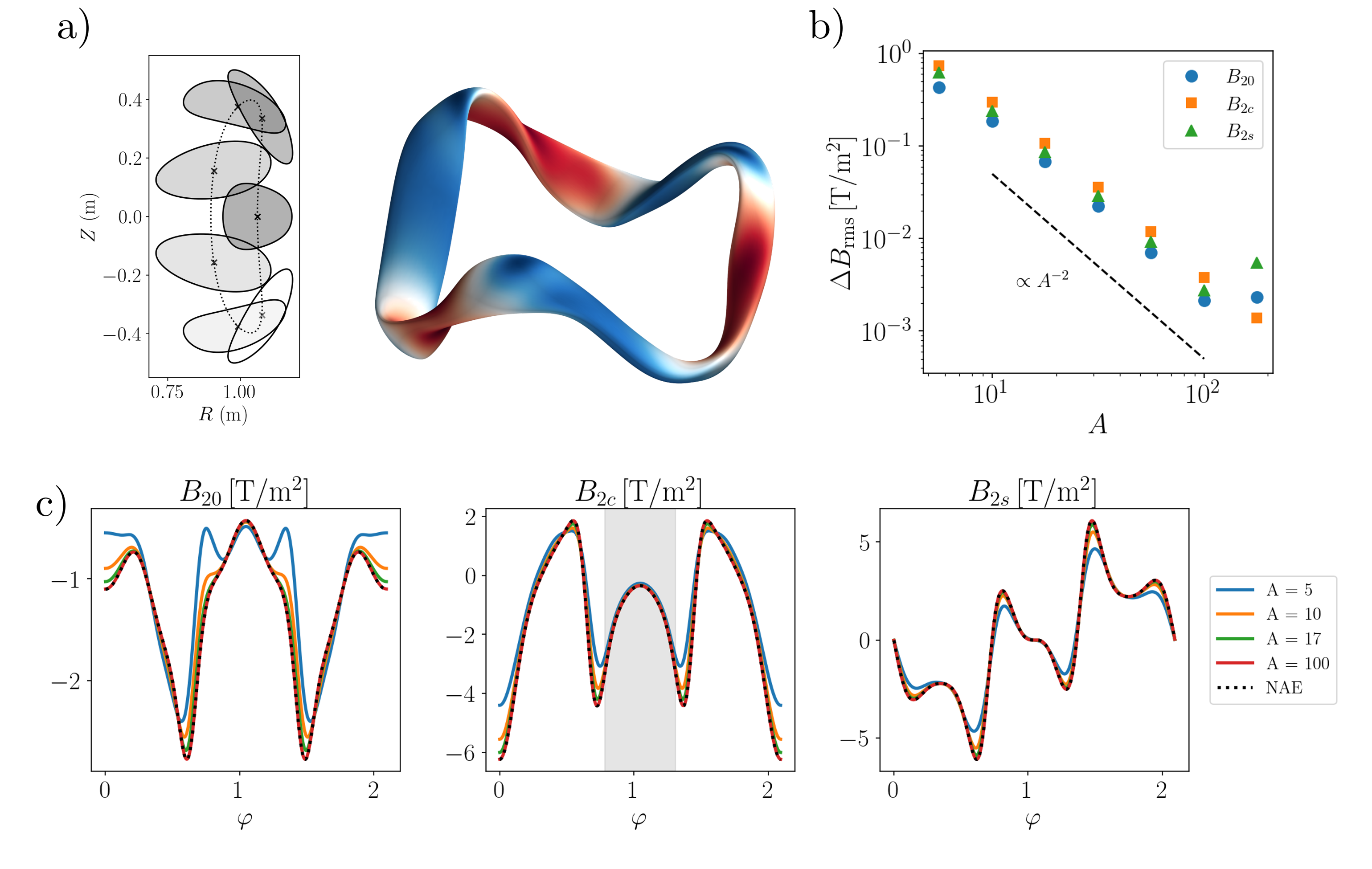}
    \caption{\textbf{Global equilibrium construction from the second-order near-axis field in Sec.~\ref{sec:opt_QI_config}.} The figure presents the second order near-axis construction based on \cite{camacho-mata-2022} presented in Sec.~\ref{sec:opt_QI_config} which has been reshaped to satisfy the QI criterion at second order near $B_\mathrm{min}$. (a) Cross-sections at constant cylindrical angle in a half period and 3D rendering at $A=10$. The dotted line traces the magnetic axis, with crosses representing the intersection with the cross-sections. (b) Root-mean-square difference in the second order near-axis magnetic field between the ideal near-axis description and the finite aspect ratio global equilibrium construction with \texttt{VMEC}. A reference quadratic scaling $\propto 1/A^2$ is given, from which the convergence of the error deviated at larger aspect ratios. (c) Explicit comparison between the different poloidal components of the second order magnetic field magnitude from the global \texttt{VMEC} equilibrium at different aspect ratios, and the ideal near-axis value (black dotted line). The shaded region indicataes the $\varphi$ range optimised for QI at second order.}
    \label{fig:NFP3_Katia_opt}
\end{figure}
Once again, the comparison to \texttt{VMEC} shows agreement (see Figure~\ref{fig:NFP3_Katia_opt}) over a range of aspect ratios, struggling to converge at very large aspect ratios, likely driven by the high degree of toroidal variation.

\section{Conclusions}
In this paper, we present the details and basis for the construction of near-axis QI equilibrium fields to second order. We describe and provide a physical guide into the construction of these fields to second order, including a careful treatment of some aspects of the problem that are intrinsic to the QI problem and deviate significantly from the previously studied quasisymmetric scenario. In particular we analyse continuity and smoothness questions as well as the special considerations required to handle the class of half helicity configurations, of which a more thorough description is provided. 
\par
The interaction of equilibrium with omnigeneity and the poloidal topology of $|\mathbf{B}|$ contours is explicitly explored. We show how the construction at first order directly impacts the behaviour at second order, especially near the field extrema. With that we learn that only special choices of these parameters and functions are consistent with a high degree of omnigeneity.
\par
We present a number of numerical near-axis constructions that demonstrate the correctness of the second order near axis construction and its asymptotic agreement with global equilibria solved with \texttt{VMEC}. This paper is instrumental in setting a solid basis off which to extend our understanding of QI fields (second order opens the door to MHD stability, particle precession, etc.). Second order may thus be used to assess equilibria within the near-axis framework, but also potentially, to provide refined versions of near-axis fields to aid in stellarator optimisation.

\section*{Data availability}
The data and scripts that support the findings of this study are openly available at the Zenodo repository with DOI/URL 10.5281/zenodo.13860071.

\section*{Acknowledgements}
The authors would like to thank the help provided by J. Geiger, S. Lazerson and M. Cianciosa with \texttt{VMEC}, as well as fruitful discussion with M. Landreman and D. Panici. 

 \section*{Funding}
E. R. was supported by a grant by Alexander-von-Humboldt-Stiftung, Bonn, Germany, through a postdoctoral research fellowship. 

\appendix 
\section{Theory of stellarator symmetric space curves with flattening points} \label{app:ss_space curves}
Let us start this appendix by setting our object of study: a regular closed curve $\gamma:~S^1\rightarrow\mathbb{R}^3$. By regular we mean that the curve is continuous and infinitely differentiable, parametrised by the length $\ell$ along the curve. The curve length is defined to be $L$, which is the period of the parametrisation. Such a curve may be described by its $\mathbb{R}^3$ embedding $\mathbf{r}_\mathrm{axis}(\ell)$.
\par
In order to introduce the standard notion of stellarator symmetry \citep{dewar1998stellarator}, we need to frame the curve in a cylindrical coordinate system. In particular, we specialise for simplicity to curves that may be parametrised by the cylindrical angle $\phi$ in such a way that $\ell(\phi)$ is a bounded, strictly monotonically increasing function of $\phi$ with $\ell(0)=0$ and $\ell(2\pi)=L$. This guarantees a one-to-one between $\ell$ and $\phi$, which we shall also take to be $C^\infty$. The curve is therefore $2\pi$-periodic in $\phi$. This consideration rules out special curve shapes such as those whose projection onto the $z=0$ plane do not correspond to the boundary of a \textit{star-shaped domain} or curves with straight sections in the $z$ direction. With this, we write our embedded curve as
\begin{equation}
    \mathbf{r}_\mathrm{axis}(\phi)=R(\phi)\hat{\pmb R}(\phi)+Z(\phi)\hat{\pmb z}, \\
\end{equation}
where $\{\hat{\pmb R},\hat{\pmb \phi}, \hat{ \pmb z}\}$ represent the orthonormal basis of the cylindrical coordinates. We consider the functions $R$ and $Z$ to be $C^\infty$ and $2\pi$-periodic in $\phi$, to provide a regular, smooth curve. The introduction of the cylindrical coordinate system might appear needlessly intrusive, but it is necessary to define stellarator symmetry. 
\par
We say a curve to be stellarator symmetric \citep{dewar1998stellarator} about a point $\mathbf{r}_\mathrm{axis}(\phi_0)$ if $R(\phi_0+\hat{\phi})=R(\phi_0-\hat{\phi})$ and $Z(\phi_0+\hat{\phi})=-Z(\phi_0-\hat{\phi})$. That is, $R$ is even and $Z$ is odd about \textit{stellarator-symmetric points}. Note that by virtue of periodicity, stellarator symmetric points always come in pairs: $(\phi_0,\phi_0+\pi)$ for $0\leq\phi_0<\pi$. 
\par
With our curve defined, we may try to construct the tangent to the curve, $ \hat{\pmb t}$. By definition,
\begin{equation}
    \hat{\pmb t}=\frac{\mathrm{d}\mathbf{r}_\mathrm{axis}}{\mathrm{d}\ell}=\frac{1}{\ell'}\left[R'\hat{\pmb R}+R\hat{\pmb \phi}+Z'\hat{\pmb z}\right] \; : \; \begin{pmatrix} \mathrm{O} \\ \mathrm{E} \\ \mathrm{E} \end{pmatrix}_\mathrm{cyl},
\end{equation}
where $\ell'=\sqrt{R^2+(R')^2+(Z')^2}>0$ and primes denote derivatives with respect to $\phi$  (not $\varphi$ as in the main text). The last part of the equality indicates the parity about the stellarator symmetric point that the three components of the vector (in the cylindrical basis) have following the properties of $R$ and $Z$; namely, even (E) and odd (O). This means that at stellarator symmetric points the curve must be orthogonal to $\hat{\pmb R}$.
\par
To understand how this curve twists, we shall continue taking additional derivatives of the curve. Constructing,
\begin{equation}
    \mathbf{r}_\mathrm{axis}''=\left(R''-R\right)\hat{\pmb R}+2R'\hat{\pmb \phi}+Z''\hat{\pmb z},
\end{equation}
we define the vector orthogonal to the curve,
\begin{equation}
    \mathbf{v}=\mathbf{r}_\mathrm{axis}'\times\mathbf{r}_\mathrm{axis}''=\begin{pmatrix} R' \\ R \\ Z' \end{pmatrix}_\mathrm{cyl}\times\begin{pmatrix} R''-R \\ 2R' \\ Z'' \end{pmatrix}_\mathrm{cyl}
    =\begin{pmatrix} RZ''-2R'Z' \\ Z'(R''-R)-R'Z'' \\ 2(R')^2+R(R-R'') \end{pmatrix}_\mathrm{cyl} \; : \;
    \begin{pmatrix} \mathrm{O} \\ \mathrm{E} \\ \mathrm{E} \end{pmatrix}_\mathrm{cyl}. \label{eqn:aux_cross_r}
\end{equation}
This vector $\mathbf{v}$ has a very special interpretation in the context of space curves, as it is parallel to the conventional Frenet binormal vector \citep{animov2001differential,mathews1964mathematical}, which we will denote $\hat{\pmb \tau}_\text{F}=\mathbf{v}/|\mathbf{v}|$. The binormal is uniquely defined then provided $\mathbf{r}_\mathrm{axis}'$ and $\mathbf{r}_\mathrm{axis}''$ are linearly independent (and $\mathbf{r}_\mathrm{axis}''$ is non-vanishing). 
\par
In addition to the above, the vector $\mathbf{v}$ can also be directly linked to the notion of curvature, $\kappa$, defined as 
\begin{equation}
    |\kappa|=\frac{1}{(\ell')^3}|\mathbf{v}|=\frac{1}{(\ell')^3}\sqrt{\left(v^R\right)^2+\left(v^\phi\right)^2+\left(v^z\right)^2},
\end{equation}
where we have written the components of $\mathbf{v}$ explicitly. 
\par
We call a point $\mathbf{r}_\mathrm{axis}(\phi_0)$ at which $\kappa=0$ a \textit{flattening point}. We characterise such points by a natural number $\nu\in\mathbb{N}_{>0}$ which we call the \textit{order} of the flattening point or of the zero of curvature, defined as the order of the first non-vanishing $\phi$ derivative of the curvature $\kappa$. The properties of this point will determine the behaviour of the binormal in its neighbourhood.
\par
Let us now specialise to curves for which every point of stellarator symmetry is also a flattening point, as is required by omnigenity at first order. Because of the well defined parity of the components of $\mathbf{v}$, it is straightforward to see that the flattening point condition requires $v^\phi(\phi_0)=0=v^z(\phi_0)$ (which are the even components). If the curvature were non zero at $\phi_0$, then the binormal would be perpendicular to the radial direction, but this is not necessarily the case when $\kappa=0$, because the leading components in the $\phi$ and $z$ directions also vanish. In fact, the orientation of the binormal in the neighbourhood of the point (and we say neighbourhood, because at the flattening point $\mathbf{v}=0$) depends critically on which the order of the zero is. It is convenient to rewrite the order $\nu$ for a stellarator symmetric flattening point as
\begin{equation}
    \nu=\min\left[ \{n\in\mathbb{N}^\mathrm{even} : \partial_\phi^n v^\phi \neq 0~\mathrm{or}~\partial_\phi^n v^z \neq 0\} \cup \{n\in\mathbb{N}^\mathrm{odd} : \partial_\phi^n v^R \neq 0\}\right],
\end{equation}
where parity of the various terms has been here leveraged. Depending on whether $\nu$ is even or odd, it is the odd or even parity components of $\mathbf{v}$ respectively that will be first non-zero in the neighbourhood of the flattening point. This leading component defines the direction of $\hat{\pmb \tau}_\text{F}=\mathbf{v}/|\mathbf{v}|$. If $\nu\in\mathbb{N}^\mathrm{even}$, the binormal is orthogonal to $\hat{\pmb R}$, $\hat{\pmb \tau}_\text{F}\cdot \hat{\pmb R}=0$. Of course, as all the components of the vector $\mathbf{v}$ are $C^\infty$, so will the Frenet frame across said point. In the case of an odd order $\nu$, $\hat{\pmb \tau}_\text{F}$ becomes aligned with the radial direction $\hat{\pmb R}$ near the flattening point, and does actually undergoes a flip across it. This is the case because the dominant behaviour of $\mathbf{v}$ is odd in this scenario, and thus $\mathbf{v}/|\mathbf{v}|$ points in opposite directions. The Frenet frame is therefore discontinuous across odd flattening points. 
\par
This discontinuity can be nevertheless reverted by redefining the binormal as part of the so-called signed frame \citep{plunk2019direct}, called also the $\beta$ frame by \cite{carroll2013improving}. In such a frame, the direction of the binormal is flipped across odd flattening points in such a way that it becomes continuous. The key insight is that this modified binormal can be shown to satisfy the Frenet-Serret equations, provided the curvature (and also the normal vector) are also signed. This flip makes the frame continuous across the flattening point, but it also makes it smooth.
Throughout the text we denote the signed Frenet frame as $(\hat{\pmb t}, \hat{\pmb \kappa},\hat{\pmb \tau})$, distinct from the conventional Frenet frame, $(\hat{\pmb t}, \hat{\pmb \kappa}_\text{F},\hat{\pmb \tau}_\text{F})$.  Likewise we reserve the symbol $\kappa$ to denote the signed curvature.  Upon flipping the sign about an odd zero, 
\begin{equation}
    \hat{\pmb \tau}(\phi) \; : \; \begin{pmatrix} \mathrm{E}^o \\ \mathrm{O}^o \\ \mathrm{O}^o \end{pmatrix}_\mathrm{cyl}, \label{eqn:binormal_parity}
\end{equation}
where the superscript $o$ indicate the odd order of the point. For an even flattening point, these components have the opposite parity. 
\par
Now that we know how the binormal vector behaves, we can construct a complete picture of the signed Frenet-Serret frame, with the normal vector defined as
\begin{equation}
    \hat{\pmb \kappa}(\phi)=\hat{\pmb \tau}(\phi) \times \hat{\pmb t}(\phi) \; : \; \begin{pmatrix} \mathrm{O}^o \\ \mathrm{E}^o \\ \mathrm{E}^o \end{pmatrix}_\mathrm{cyl}. \label{eqn:normal_parity}
\end{equation}
Just to be clear, if the order of curvature is even, then the parity of the components of $\hat{\pmb \kappa}$ and $\hat{\pmb \tau}$ are reversed.
\par
There is an interesting question to be asked here about whether the signed frame is continuous or not globally. We have argued about these properties locally across a flattening point, but after going around the whole curve, the question remains of whether after all the flipping involved, the frame at $\phi=0$ is the same as $\phi=2\pi$. To that end, let us label each of the flattening points in our curve by an index $i\in\mathbb{N}: 0\leq i<n$, where $n\in\mathbb{N}^\mathrm{even}$ is the total number of points. To each point we assign an order of zeros $\nu_i$. As we follow the axis, we define the total order of the curve as $\mathcal{V}=\sum_{i=0}^{n-1}\nu_i$. The signed frame of a curve with an even order will be globally continuous. This is obviously true if all flattening points are even order; in case of an even number of odd such points, every time we cross one such point we gain a sign flip, and thus if an even number of these flips is to be counted, after a complete period, we are back to the start. However, if $\mathcal{V}\in\mathbb{N}^\mathrm{odd}$, we are left with a frame that is discontinuous at the edges of the period. The signed frame has a sign flip there. If in some abstract sense we were to extend the curve for another period, the frame would come back to itself. The signed frame for odd $\mathcal{V}$ curves is $4\pi$-periodic.
\par
The continuity of the frame within the period allows us to define the notion of \textit{helicity} of the axis. We define this in the way that the \textit{self-linking number} is defined in a curve free from flattening points \citep{moffatt1992helicity,fuller1999geometric,fuller1971writhing,oberti2016torus,rodriguez2022phases}. It is the linking number between the axis $\mathbf{r}_\mathrm{axis}$ and the curve generated by an infinitesimal displacement of the curve in the direction of the \textit{signed} binormal vector, $\mathbf{r}_b^\mathcal{\delta}(\phi)=\mathbf{r}_\mathrm{axis}(\phi)+\delta~\hat{\pmb \tau}(\phi)$. Note that in the case of even $\mathcal{V}$ curves this definition needs no further comment. In the case of odd $\mathcal{V}$ we need to be more careful as $\mathbf{r}_b^\delta$ is not a closed curve in a single $\phi$ period, and thus we need to define that curve using two toroidal transits, for which the function $\hat{\pmb \tau}$ is naturally extended to be $4\pi$-periodic. With that, we define helicity in that case as the linking number between those two closed curves but divided by two. This division by two leads to the notion of \textit{half period} curves. In practice, computing the helicity of a curve does not require the explicit construction of the auxiliary curve $\mathbf{r}_b^\mathcal{\delta}$, and  can be computed by counting the integer number of times that the binormal vector performs complete rotations about the axis in the $(R,z)$ plane \citep{landreman2019,rodriguez2022phases}. 
\par
Let us now specialise to the case of having only two distinct flattening points within a field period. This is the typical near-axis QI scenario in which there is a single magnetic well per period \citep{parra2015less}. As stellarator symmetric points come in pairs (see above), we put them at the centre and edge of the domain. We take the central point (where the bottom of the well is) to be of odd order, as required for QI \citep{camacho-mata-2022, rodriguez2023higher}. In that case, an odd order maximum leads to an even $\mathcal{V}$ curve, an integer helicity and a continuous frame. That helicity ends up being an integer can also be argued geometrically by observing the behaviour of the normal at the flattening points. At both points $\hat{\pmb \tau}\propto\hat{\pmb R}$, and thus by periodicity the number of turns of the binormal has to be an integer.  The case of half integer helicity arises with an even flattening point at the edge of the domain. In this case $\hat{\pmb \kappa}\cdot\hat{\pmb R}$ is non-zero at such locations, and by parity, the normal must have opposite directions at each edge. This leads to a total of an odd number of $\pi$ rotations, {\it i.e.} the half-helicity scenario.

\section{Continuity of the field} \label{app:cont_field}
\subsection{Continuity of half-helicity fields} \label{app:cont_half_period}
Let us start our discussion of the continuity of the near-axis field construction by looking at the inverse-coordinate description of flux surfaces. We write in real space,
    \begin{equation}
        \mathbf{r}=\mathbf{r}_\mathrm{axis}+X\hat{\pmb \kappa}+Y\hat{\pmb \tau}+Z\hat{\pmb t}, \tag{\ref{eqn:inverse-coordiantes_FS}}
\end{equation}
where $\{\hat{\pmb t},\hat{\pmb \kappa}, \hat{\pmb \tau}\}$ represent the signed Frenet triad of a regular, smooth, stellarator symmetric axis as described in Appendix~\ref{app:ss_space curves}. We shall consider the axis to have $N$-fold symmetry, and each period to have two flattening points, which coincide with a pair of stellarator symmetry points. In addition, choose the order of the mid point to be odd (which will represent in the near-axis description the point of minimum $|\mathbf{B}|$ along the axis) and label it as our origin $\varphi=0$. 
\par
\subsubsection{Integer helicity axes}
The case of an integer helicity axis (equivalently, an even $\mathcal{V}$ curve) is straightforward. First, regarding its construction, we showed in Appendix~\ref{app:ss_space curves} that this requires the order of the flattening at the edge of the period to be odd. In such a scenario the signed Frenet frame is continuous and smooth, meaning that so will $\mathbf{r}$, Eq.~(\ref{eqn:inverse-coordiantes_FS}), provided the functions $X,~Y$ and $Z$ are themselves $C^\infty$, and periodic functions in $\varphi$ and $\theta$. Thus the near-axis framework can be set-up in the context of periodic, smooth functions. This scenario is the same as it is for quasisymmetric fields \citep{landreman2019}.

\subsubsection{Half helicity axes}
The half helicity scenario is more subtle. Its construction requires a flattening point of even order at the edge of the domain, which, as discussed in the previous section, implies that the fractional rotation of the signed Frenet over the field period leaves it flipped.  More precisely, this means that for $\varphi \in[0,2\pi/N)$,
\begin{align}
    \hat{\pmb \kappa}^i(\varphi)=-\hat{\pmb \kappa}^i(\varphi+2\pi/N),~~ \hat{\pmb \tau}^i(\varphi)=-\hat{\pmb \tau}^i(\varphi+2\pi/N), \nonumber\\
    \hat{\pmb t}^i(\varphi)=\hat{\pmb t}^i(\varphi+2\pi/N),~~ \kappa(\varphi)=-\kappa(\varphi+2\pi/N),
\label{eqn:half_period_FS}
\end{align}
where the superscripts denote the cylindrical components of the basis vectors.  If it was not for the minus sign, then the frame would be continuous and $2\pi/N$-periodic, but we see for the half-helicity case that it behaves as if its period was not $2\pi/N$, but rather $4\pi/N$. That is, the signed Frenet frame becomes `half-periodic' (i.e. periodic in the extended double-period domain). 
\par
A function that is smooth and continuous in the extended domain but has a sign jump in the real domain can generally be written as,
\begin{equation}
    f(\varphi)=-f(\varphi+2\pi/N) \Leftrightarrow f(\varphi)=\tilde{f}_c(\varphi)\cos(N\varphi/2)+\tilde{f}_s(\varphi)\sin(N\varphi/2), \label{eqn:half_periodic_functions}
\end{equation}
where $\tilde{f}_{c/s}$ are smooth and periodic in $2\pi/N$. This form follows from consideration of a function $f=\cos a\varphi$ and applying the condition that $f(\varphi)=-f(\varphi+2\pi/N)$, which implies $a=N/2+nN$ where $n\in\mathbb{Z}$. 
\par
An interesting property of these half periodic functions is the following: any product of two half periodic functions, Eq.~(\ref{eqn:half_periodic_functions}), is a smooth periodic function. Therefore, in order for $\mathbf{r}$, Eq.~(\ref{eqn:inverse-coordiantes_FS}), to be smooth considering Eq.~(\ref{eqn:half_period_FS}), we need the functions $X$ and $Y$ to be half periodic functions.

\subsubsection{Continuity of first order construction for half helicity}
Let us consider the first order construction in the context of the requirements above. At this order, see Fig.~\ref{fig:1st_order_nae_equilibrium}, the functions $X_1$ and $B_1$ are the first to appear. We have just argued that the former must be a half-period function, while the latter should not because it is a physical quantity. Following \cite{garrenboozer1991a, plunk2019direct}, and in particular the prescription in \cite{Camacho2023helicity} we write,
\begin{equation}
    X_1=\bar{d}(\varphi)\cos\left[\chi-\alpha_1(\varphi)\right],
\end{equation}
where $\chi=\theta-M\varphi$, and $M$ is the half-helicity of the signed frame. As shown in Section~5 of \cite{Camacho2023helicity}, the periodicity of $\alpha_1$ and $\bar{d}$ are sufficient to ensure that $X_1(-\pi/N)=-X_1(\pi/N)$. Here we can go further and claim this condition for all derivatives, simply by recognising that as written, $X_1$ is a half-periodic function, Eq.~(\ref{eqn:half_periodic_functions}).  
\par
Now, we must make sure that the magnetic field magnitude is continuous and periodic. From Eq.~(A22) in \cite{landreman2019}, 
\begin{equation}
    B_1=\kappa X_1 B_0,
\end{equation}
where we recall that $\kappa$ is the signed curvature, which is also a half-periodic function.  We have already seen that the product of two half-period functions is periodic, and thus so is $B_1$.
\par
To complete the picture at first order (see Fig.~\ref{fig:1st_order_nae_equilibrium}) it suffices to check that the same argument as for $X_1$ holds for $Y_1$.\footnote{In this case, we have an additional function $\sigma(\varphi)$ involved. This is the solution to the $\sigma$-equation, Eq.~(\ref{eqn:sigma_eqn}), which sees no half-periodic functions explicitly at all.} Thus, the prescription of \cite{Camacho2023helicity} does indeed guarantee the correct behaviour of the near-axis construction to first order. Like for quasisymmetric stellarators \citep{landreman2019, rodriguez2022phases}, the helical angle $\chi$ is particularly convenient to use as "poloidal" angle respect to which the harmonic content in the near-axis expansion is defined. For the half-periodic $X_1$ and $Y_1$ this means that $X_{1c}$, $Y_{1s}$, etc. are periodic in the near-axis equilibrium equations of \cite{landreman2019}. This proves convenient when proceeding to second order in the expansion as well.
\par
Before moving to next order, we summarise the periodicity and parity of the first order functions in Table~\ref{tab:firs_order_parity}.
\begin{table}
    \centering
    \begin{tabular}{ccc||ccc}
         & Parity & Period~ & ~Period & Parity & \\\hline
        $X_{1c}$ & O & $2\pi/N$ & $2\pi/N$ & E & $X_{1s}$ \\
        $B_{1c}$ & E$^o$ & $4\pi/N$ & $4\pi/N$ & O$^o$ & $B_{1s}$ \\
        $Y_{1c}$ & E & $2\pi/N$ & $2\pi/N$ & O & $Y_{1s}$ \\
    \end{tabular}
    \caption{\textbf{Periodicity and parity of first order near-axis.} Table summarising the periodicity and parity of functions involved in the first order near-axis expansion. The parity is here considered respect to the centre of the domain (the bottom of the well), with the superscript $o$ indicating that the parity applies to the point of odd curvature; note that the opposite parity applies at the edges of the domain in a half-helicity scenario. The functions correspond to the $\chi$-harmonics of $X_1$, $Y_1$ and $B_1$. }
    \label{tab:firs_order_parity}
\end{table}

\subsubsection{Continuity of second order construction: half helicity}
Let us continue the construction to second order. Following the flow of the construction in Fig.~\ref{fig:2nd_order_nae_equilibrium}, the first element to consider is $B_\psi$ which is directly connected to $B$ and $p$ through MHS equilibrium. Although these are physical quantities, resolving the function in harmonics of the newly defined $\chi=\theta-M\varphi$ (see first order discussion) makes functions $B_{1c}$ and $B_{1s}$ half-periodic (see Table~\ref{tab:firs_order_parity}). Thus so will the harmonics of $B_\psi$. 
\par
To construct the function $Z$ next we only need first order periodic quantities, with which all harmonics of $Z$ are smooth and periodic. Note that resolving $Z$ in $\chi$-harmonics does not introduce the half-periodicity issues of $B_{\psi1}$ or $B_1$. The reason is that at second order, harmonics involve $2\chi=2\theta-2M\varphi$, which becomes periodic under a sine or cosine. This points to a potential issue that arises at second order with the other elements, $X_2$ and $Y_2$, describing flux surfaces. Unlike at first order, the convenient definition of $\alpha_1$ cannot account for the necessary half-periodicity of the harmonics. We must impose the half-periodic behaviour more explicitly.
\par
Consider $X_{2c}$ and $X_{2s}$, which are inputs to the near-axis construction (see Figure~\ref{fig:2nd_order_nae_equilibrium}). We must choose these to be half period functions, but they must also have a particular parity. Through Eqs.~(A34-36) of \cite{landreman2019}, $X_2$ components are directly linked to $B_2$, which must satisfy $B(\chi,\varphi)=B(-\chi,-\varphi)$ by virtue of stellarator symmetry. Therefore $B_{2c}$ and $B_{2s}$ must be even and odd functions of $\varphi$ respectively. As $\ell'\kappa X_{2c}=B_{2c}+\hat{\mathcal{T}}_c$, and $\hat{\mathcal{T}}_c$ can be shown to be even, it must be the case that $X_{2c}$ is odd. Odd parity in addition to its half-periodic nature means that we may generally provide $X_{2c}=\tilde{X}_{2c}^o\cos(N\varphi/2)+\tilde{X}_{2s}^e\sin(N\varphi/2)$, where the tilde functions are periodic well defined parity (e-even or o-odd) of $\varphi$. The product through $\kappa$, as it occurred at first order, guarantees the consistency of $X$ half-period condition with the periodic nature of $B$. A similar argument follows for $X_{2s}$ and $B_{2s}$, but with the parity reversed; i.e. $X_{2s}$ must be even and $B_{2s}$ odd. 
\par
The next elements in the construction are $X_{20}$ and $Y_{20}$, which are the solutions to the system of equations comprising Eqs.~(A41-48) in \cite{landreman2019}. Term by term, one can be assured that the solution to the system of equations will be a half-period function provided that the prescription so far followed is continued. This means that we cannot exploit periodicity in solving this system of coupled ODEs. In particular, if we attempt a solution of the system of ODEs as the solution of the linear system $(\pmb{D}+\pmb{A})\mathbf{x}=\mathbf{c}$ where $\mathbf{x}$ contains the values of functions $X_{20}$ and $Y_{20}$ on regularly spaced collocation points, we cannot use pseudo-spectral matrices \citep{weideman2000matlab} like in \cite{landreman2019} to solve the problem. We can nevertheless extend the treatment by defining the following pseudo-spectral differentiation matrix for equally spaced grid points of a half-periodic function,
    \begin{equation}
        {\pmb D}_{ij}=\begin{cases}
            \begin{aligned}
                &0, & (i=j) \\
                &\frac{N}{2}(-1)^{i-j}\cot\left[\frac{(i-j)h}{2}\right], & (i\neq j,~n\in\mathbb{N}^{\mathrm{odd}})  \\
                &\frac{N}{2}(-1)^{i-j}\csc\left[\frac{(i-j)h}{2}\right], & (i\neq j,~n\in\mathbb{N}^{\mathrm{even}})
            \end{aligned}
        \end{cases}
    \end{equation}
for $n$ the number of evenly spaced grid points and $h=2\pi/n$.
\par
Once the system of equations is solved using this correct definition of $\pmb{D}$, the final step in the second order construction (see Figure~\ref{fig:2nd_order_nae_equilibrium}) is straightforward. The algebraic operations can be seen to preserve the right half-periodic behaviour of $Y_{2c}$ and $Y_{2s}$, while preserving the periodic nature of $B_{20}$ through multiplication by $\kappa$. We have therefore shown that a consistent construction to second order is possible for half-helicity fields. We summarise the periodicity and parity of the second order functions in Table~\ref{tab:second_order_parity}.
\par
A final comment should be devoted to the scenario at third order. We only need third order elements when it comes to constructing a description of the equilibrium boundary when the near-axis is used to construct global equilibria. This is important in Section~\ref{sec:numerical_bench}, the numerical comparison. The details and procedure can be found in Section~3 of \cite{landreman2019}, and here we must make sure that the shapes constructed following that are continuous and smooth. The answer is affirmative, as it is only the $\chi$ (and not $3\chi$) harmonics that are needed to be non-zero, and the same way that the first order $X_1$ and $Y_1$ were well behaved, so are these. The procedure needs no change to accommodate the half-helicity problem.

\begin{table}
    \centering
    \begin{tabular}{ccc||ccc}
         & Parity & Period~ & ~Period & Parity & \\\hline
        $B_{\psi1c}$ & O$^o$ & $4\pi/N$ & $4\pi/N$ & E$^o$ & $B_{\psi1s}$ \\
        $X_{2c}$ & O$^o$ & $4\pi/N$ & $4\pi/N$ & E$^o$ & $X_{2s}$ \\
        $B_{2c}$ & E & $2\pi/N$ & $2\pi/N$ & O & $B_{2s}$ \\
        $Y_{2c}$ & E$^o$ & $4\pi/N$ & $4\pi/N$ & O$^o$ & $Y_{2s}$ \\
        $X_{20}$ & O$^o$ & $4\pi/N$ & $4\pi/N$ & E$^o$ & $Y_{20}$ \\
        $Z_{2c}$ & O & $2\pi/N$ & $2\pi/N$ & E & $Z_{2s}$ \\
        $Z_{20}$ & O & $2\pi/N$ & $2\pi/N$ & E & $B_{20}$ \\
    \end{tabular}
    \caption{\textbf{Periodicity and parity of second order near-axis.} Table summarising the periodicity and parity of functions involved in the second order near-axis expansion. The parity is here considered respect to the centre of the domain (the bottom of the well), with the superscript $o$ denoting the different parity across the edges of the domain in a half-helicity scenario. }
    \label{tab:second_order_parity}
\end{table}

\subsection{Continuity of buffer region} \label{app:cont_buff}
We refer as \textit{buffer regions} to the parts of the toroidal domain (which must include $B_\mathrm{max}$) in which a controlled deviation from QI is allowed for. This is a first order near-axis notion \citep{plunk2019direct,camacho-mata-2022} arising from a clash between omnigeneity and the periodicity requirement of the field. Let us see how it arises and how it can be managed in a controlled way.
\par
To this end, let us start by writing the first order field as $B_1=\bar{d}\kappa\cos(\chi-\alpha_1)$. In order to make the field periodic, we define here $\chi=\theta-M\varphi$ with $M$ being the helicity of the axis (which we may take to be an integer for simplicity), and $\bar{d}$ and $\alpha_1$ as periodic functions.\footnote{It may appear sufficient although not necessary to choose $M$ to match the helicity of the axis, as adding a term $n\varphi$ for any $n\in\mathbb{Z}$ would also fit the bill. However, the choice of the axis helicity is unique in achieving a definition for $\theta$ consistent with its poloidal, and not helical, interpretation. To see how this is the case we can follow closely the argument in \cite{landreman2018a}. Consider describing the position of the line of constant $\chi-\alpha_1$ as $\mathbf{x}(\chi,\varphi)$, and consider its projection onto $\hat{\pmb \kappa}$. By Eq.~(\ref{eqn:inverse-coordiantes_FS}), $p=\mathbf{x}\cdot\hat{\pmb\kappa}=c\bar{d}$ where $c=\cos(\chi-\alpha_1)$. Given that $c$ is a constant, let us take $c>0$ (the other cases can be shown similarly), in such a way that the sign of $p$ depends on $\bar{d}$. However, by construction $\bar{d}>0$, and thus the position of the constant $\chi-\alpha_1$ line cannot deviate more than $\pi/2$ from $\hat{\pmb \kappa}$ (as it would have to cross the zero otherwise). This means that upon a whole toroidal transit, the line of constant $\chi-\alpha_1$ is glued to the normal of the axis, and thus must complete a number of turns equal to $M$, the helicity of the axis. If $\theta$ is thus meant to have its poloidal meaning, then $\alpha_1$ should indeed be periodic. If we don't do so, $\theta$ becomes a helical angle which even modifies the value of the rotational transform as it appears in Eq.~(\ref{eqn:B_contra}). It is not that the equations are incorrect, but their interpretation would be non-standard.} The additional constraint of stellarator symmetry requires $\bar{d}$ to be even and $\alpha_1=\pi/2+\tilde{\alpha}$, where $\tilde{\alpha}$ is an odd, periodic function. Finally, we have the QI conditions \citep{plunk2019direct}\citep[Eq.~(31)]{rodriguez2023higher} which require,
\begin{equation}
    \alpha_1^\mathrm{QI}=\pi/2+\bar{\iota}\varphi. \label{eqn:ideal_QI_alpha}
\end{equation}
But clearly simply setting $\alpha_1 = \alpha_1^\mathrm{QI}$ violates periodicity, unless $\bar{\iota}=\iota-M=0$, which is generally not true. This is why it is generally not possible to grant QI exactly. For an approximate QI field, we need the choice of an odd periodic $\tilde{\alpha}$ that is close to the ideal QI behaviour, Eq.~(\ref{eqn:ideal_QI_alpha}).

\subsubsection{Piecewise buffer}
The first approach considered in the literature to construct $\tilde{\alpha}$ can be found in \cite{plunk2019direct}. The periodic function is defined as a piecewise function,
\begin{equation}
    \tilde{\alpha}=\begin{cases}
        \bar{\iota}\varphi, \quad & (|\varphi|<\pi/N-\delta) \\
        \frac{\bar{\iota}}{\delta}(\pi/N-\delta)(\pi/N-\varphi), \quad & (\pi/N\geq\varphi>\pi/N-\delta) \\
        -\tilde{\alpha}(-\varphi), \quad & (-\pi/N\leq\varphi<-\pi/N+\delta)
    \end{cases}
\end{equation}
where the central region of the domain is exactly omnigeneous, and the break-up is restricted to narrow regions of width $\delta$ at the edges (input parameter). The piecewise function is $C^0$, with the derivative being discontinuous at the joints of the domain. This lack of smoothness leads to an impossibility to proceed to second order, where the field becomes discontinuous. In addition, and as studied in \cite{camacho-mata-2022}, the edge and the core being decoupled in this way leads to large variations in the field within the edge buffer, where the field needs to make, within a narrow space, the necessary corrections for a consistent field.

\subsubsection{Standard buffer} \label{sec:extended_buffer}
Recognising the limitations of the above, in the recent work by \cite{camacho-mata-2022}, a construction was presented which attempted a more smooth distribution of the buffer. In our present notation, that construction may be written as 
    \begin{equation}
        \tilde{\alpha} = \bar{\iota}\frac{\pi}{N}\left[\frac{\varphi}{\pi/N}-\left(\frac{\varphi}{\pi/N}\right)^{2k+1}\right], \label{eqn:katia_alpha}
    \end{equation}
which is as needed an odd function of $\varphi$ with the correct QI behaviour near $\varphi\sim0$. The order $k>0$ denotes the number of derivatives ($2k+1$) that $\tilde{\alpha}$ match the ideal QI form Eq.~(\ref{eqn:ideal_QI_alpha}) near the centre of the domain. The larger $k$ is, the larger the central region of agreement, and thus the narrower the region with a significant deviation at the edges. To avoid some of the problems of the piecewise construction, typical values of $k=2$ or $3$ are appropriate  \citep{camacho-mata-2022,Camacho2023helicity}. We call this construction the \textit{standard buffer}.
\par
Although the function is not piecewise, it is not smooth in the periodic domain either. The lack of smoothness is localised at the edges, $\varphi=\pm\pi/N$, where $\cos\tilde{\alpha}$ and $\sin\tilde{\alpha}$ are $C^2$ and $C^1$ respectively. This means that first order field quantities such as $B_1$ or $X_1$ will be clearly non-smooth.\footnote{In fact $X_{1s}$ is $C^2$ and $X_{1s}$, $Y_{1s}$, $Y_{1c}$, $\sigma$ and $B_1$ are $C^1$.} In fact, the lack of smoothness is so readily observed that second order field quantities will be unphysically discontinuous. Let us show this briefly.
\par
Consider explicitly the evaluation of $B_{2s}$, which directly depends on the derivative of $Z_{2s}$, and as such, see Fig.~\ref{fig:2nd_order_nae_equilibrium}, involves second derivatives of first order quantities as
\begin{equation}
    Z_{2s}=-\frac{1}{8\ell'}(V_2'-2\bar{\iota}V_3),
\end{equation}
where
\begin{subequations}
    \begin{align}
        V_2&=2[X_{1c}X_{1s}+Y_{1c}Y_{1s}], \\
        V_3=&X_{1c}^2-X_{1s}^2+Y_{1c}^2-Y_{1s}^2.
    \end{align}
\end{subequations}
It can be shown through the explicit calculation of the dependence on $\tilde{\alpha}$ that $V_3\in C^2$ and $V_2\in C^1$, so that $Z_{2s}\in C^0$, i.e. just about continuous. From Eq.~(A35) in \cite{landreman2019}, one may relate $B_{2s}\sim Z_{2s}'$, meaning that the function $B_{2s}$ is generally discontinuous (and non-symmetric) at the endpoints of the periodic domain. See Figure~\ref{fig:NFP2_Katia_std_buffer} for an example of this. A similar argument shows that the other components of $B$ are continuous, as well as the $X_2$ ad $Y_2$ components.\footnote{The case of $X_{20}$ and $Y_{20}$ are special, as they come from solving a set of ODEs. As such, we expect the order of differentiability of these functions to be raised by one respect to the minimally differentiable function in the system, which is $C^0$. Thus we expect $X_{20}\in C^1$ at least.} Although the discontinuity limits the physicality of the second order field, it occurs in a rather controlled fashion, as can also be seen in the example of Figure~\ref{fig:NFP2_Katia_std_buffer}. 
\par
The lack of smoothness can however limit some of the numerical procedures in the problem, such as the use of pseudo-spectral methods to solve ODEs, which rely on the assumption of periodicity and smoothness. In that case, even adopting the upgraded version of this buffer also presented in \cite{camacho-mata-2022}, which raises the order of differentiability further, would be potentially problematic. It is typical to expect Gibb's-like phenomena.

\subsubsection{Smooth buffer}
Following the previous discussion, it is natural to seek the construction of a buffer that is smooth everywhere and yet similar to  functions like that of Eq.~(\ref{eqn:katia_alpha}). Generally, we shall write $\tilde{\alpha}$, which must be an odd, periodic, smooth function, as a sine Fourier series.
\par
Following the construction of the standard buffer, Section~\ref{sec:extended_buffer}, we impose the following $k$ constraints that attempt to match $\tilde{\alpha}$ to its ideal QI form near $\varphi=0$,
\begin{equation}
        \begin{cases}
            \begin{aligned}
                &\frac{\partial\tilde{\alpha}}{\partial\varphi}=\bar{\iota}, \\
                &\left(\frac{\partial}{\partial\varphi}\right)^{2j+1}\tilde{\alpha}=0,\quad (1<j\leq k)
            \end{aligned} 
        \end{cases}\label{eqn:buff_smooth_def}
    \end{equation}
and everything is evaluated at $\varphi=0$. The larger $k$, the larger the number of derivatives near the origin that match the ideal QI case. 
\par
To satisfy these $k$ constraints uniquely, we restrict the sine Fourier series to the first $k$ harmonics, so that the conditions above reduce to a linear system of algebraic equations. Defining the vector $\mathbf{a}$ as the 1-indexed vector of Fourier coefficients, we may write
\begin{equation}
    \mathbf{a}_i=\frac{\bar{\iota}}{N}{\pmb S}^{-1}_{i1},
\end{equation}
where the matrix,
\begin{equation}
    {\pmb S}_{ij}=(-1)^{i+1}j^{2i+1}.
\end{equation}
That is, the first column of the inverse of ${\pmb S}$. At large $k$ (for $k>9$ in practice) the inversion of the matrix becomes problematic numerically due to the large discrepancy in magnitude in the matrix elements due to the strong exponential term.
\par 
Given the similarities of this buffer construction with the standard one, Section~\ref{sec:extended_buffer}, we may ask how they compare. For this comparison, we define an effective size of the buffer region as the fraction of the domain where the QI deviation is significant. In particular, we take the turning points where $\tilde{\alpha}^\prime = 0$ as measure. The comparison is shown in Table~\ref{tab:compare_buffer}.
    \begin{table}
        \centering
        \begin{tabular}{|c|cccccccc|}\hline
            $k$ & 1 & 2 & 3 & 4 & 5 & 6 & 7 & 8 \\\hline
            Smooth & 0.5 & \cellcolor[gray]{.9} 0.43 & 0.38 & 0.35 & \cellcolor[gray]{.85} 0.33 & 0.31 & 0.30 & \cellcolor[gray]{.8} 0.28 \\
            Standard & \cellcolor[gray]{.9} 0.42 & \cellcolor[gray]{.85} 0.33 & \cellcolor[gray]{.8} 0.28 & 0.24 & 0.21 & 0.19 & 0.18 & 0.16 \\\hline
        \end{tabular}
        \caption{\textbf{Effective size of the buffer region.} Table comparing the effective size of the buffer region (fraction of the domain) for different values of the parameter $k$ between the smooth and standard constructions.}
        \label{tab:compare_buffer}
    \end{table}
    The smooth buffer struggles to narrow the buffer region down, and does so at the expense of a high harmonic content. Nevertheless, following the prescription of \cite{camacho-mata-2022}, a reasonable choice of parameter would be $k=5$ for the smooth buffer (which we may also call \textit{Fourier buffer}). Other alternatives could also be concocted, but these suffice.

\section{Deviation from omnigeneity at second order} \label{app:2nd_order_deviation}

This Appendix presents an assessment of the level of omnigeneity breaking at second order in the near-axis expansion, taking into account its necessary violation at first order. An understanding of these deviations will resolve the toroidal extent about the minimum of the well in which it is meaningful to enforce the QI conditions at second order. This Appendix draws heavily on the machinery developed in \cite{rodriguez2023higher}, of which a certain level of familiarity is assumed. The most essential elements are included here, but the reader is referred to the aforementioned paper for a more thorough and complete account.
\par
Let us start by settling the type of field we are considering here: we focus on stellarator symmetric, single-well fields, of which the bottom of the well is taken to be at $\varphi=0$. We define \textit{omnigeneity} as the property of a field which satisfies,
\begin{equation}
    Y(r,\alpha,\varphi) = Y(r,\alpha,\eta(\varphi)), \label{eqn:omn_Y_cond}
\end{equation}
where
\begin{equation}
    Y=\frac{\nabla \psi\times\mathbf{B}\cdot\nabla B}{\mathbf{B}\cdot\nabla B}, \label{eqn:Ydef}
\end{equation}
and $\eta(\varphi)$ is a function that maps pairs of equal $B$ points on either side of the magnetic well along a field line; i.e. $B(\alpha,\eta(\varphi))=B(\alpha,\varphi)$ and $\eta(\varphi)\neq\varphi$ except for $\varphi=0$ and $\alpha=\theta-\iota\varphi$. This description may be interpreted as the condition for the radial magnetic drift displacement of particles to be equal but opposite on either side of the magnetic well.  
\par 
An asymptotic consideration of Eq.~(\ref{eqn:omn_Y_cond}) in the distance from the axis and careful consideration order-by-order yields the conditions of QI used in the main body of this paper and derived in \cite{rodriguez2023higher}, the lowest order of which were originally obtained in \cite{plunk2019direct}. Here, though, we are not interested on re-deriving this exact problem. Instead, we want to assess the deviation in  Eq.~(\ref{eqn:omn_Y_cond}) induced by the existing limitations to achieve omnigeneity exactly. In particular, we shall consider a field that is QI at first order but for the presence of a buffer $\alpha_\mathrm{buf}$,
\begin{equation}
    B_1=B_0(\varphi)d(\varphi)\cos\left[\alpha+\frac{\pi}{2}-\alpha_\mathrm{buf}(\varphi)\right].
\end{equation}
The field is ideally omnigeneous when $\alpha_\mathrm{buf}=0$ provided $d$ is odd ($B_0$ is even). The buffer must be an odd function of $\varphi$ in order for the field to comply with stellarator symmetry. The presence of the buffer leads to an omnigeneity breaking that we may quantify to leading order $O(r)$ as,
\begin{equation}
    \Delta Y^{(1)}=Y^{(1)}(\varphi)-Y^{(1)}(-\varphi)=2\frac{G_0B_0}{B_0'}d\sin(\alpha_\mathrm{buf})\sin\alpha.
\end{equation}
The procedure may be repeated with the next order, in this case being careful to include the variation in the map $\eta$ due to the variation of $B$. Following \cite{rodriguez2023higher}, and after significant algebra, we write using the shorthand notation $s_x=\sin x$ and $c_x=\cos x$,
\begin{align}
    \Delta Y^{(2)}=&Y^{(2)}(\varphi)-Y^{(2)}(-\varphi)-\eta^{(1)}(\varphi)\left(\partial_\varphi Y^{(1)}\right)(-\varphi)= \\
    =&-2\frac{G_0B_0}{B_0'}s_\alpha\left\{\left[4\frac{\Delta B_{2c}^\mathrm{QI}}{B_0}+\frac{1}{dB_0^2s_{\alpha_\mathrm{buf}}}\left(\frac{B_0^3d^3}{B_0'}s_{\alpha_\mathrm{buf}}^3\right)'\right]c_\alpha+\right.\nonumber\\
    &\hspace{7cm}\left.+dc_{\alpha_\mathrm{buf}}s_\alpha\left(\frac{B_0d}{B_0'}s_{\alpha_\mathrm{buf}}\right)'\right\}. \label{eqn:Y2_non_omni}
\end{align}
The shorthand $\Delta B_{2c}^\mathrm{QI}=B_{2c}^\mathrm{QI}-(B_0^2d^2/B_0')'/4$ is zero when $B_{2c}^\mathrm{QI}$ assumes its ideal QI form. Putting all together, we may write
\begin{equation}
    \Delta Y\approx 8r\frac{G_0B_0}{B_0'}s_\alpha\Biggl\{\underbrace{\frac{ds_{\alpha_\mathrm{buf}}}{4}}_{\mathcal{Y}_0}+rc_\alpha\left[\frac{\Delta B_{2c}^\mathrm{QI}}{B_0}+\mathcal{Y}_1\right]+rs_\alpha \mathcal{Y}_2\Biggr\} + O(r^3), \label{eqn:omni_breaking_2nd}
\end{equation}
where $\mathcal{Y}_i$ for $i=1,2$ may be directly read off Eq.~(\ref{eqn:Y2_non_omni}). As it must be the case following the conditions of omnigeneity, when $\alpha_\mathrm{buf}=0$, the omnigeneity condition reduces to $\Delta B_{2c}^\mathrm{QI}=0$. When we acknowledge our impossibility to satisfy QI exactly at first order in the near-axis expansion, there are multiple contributors to omnigeneity breaking at second order. Insisting on the second order QI condition, namely $\Delta B_{2c}^\mathrm{QI}=0$, would only make sense in such a scenario if the other terms in Eq.~(\ref{eqn:omni_breaking_2nd}), each with their own $\alpha$ dependence, are sufficiently small. I.e. the other terms (which solely depend on first order quantities) serve as meaningful thresholds for $\Delta B_{2c}^\mathrm{QI}/B_0$.
\par
It is reasonable to assess all of these contributions near the bottom of the well, where $\alpha_\mathrm{buf}\approx 0$, and thus we expect the QI breaking terms to be small. In fact, taking near the bottom of the well $d\sim\varphi^v$, $B_0'\sim\varphi^{u-1}$ and $\alpha_\mathrm{buf}\sim\varphi^{2k+3}$\footnote{The dependence of $\alpha_\mathrm{buf}$ on $\varphi$ near $\varphi=0$ is considering the smooth buffer as defined in Appendix~\ref{app:cont_buff}, Eq.~(\ref{eqn:buff_smooth_def}). Hence the involvement of the parameter $k$.}, we have 
\begin{equation}
    I[\mathcal{Y}_0] = 2k+3+v, \quad I[\mathcal{Y}_1] = 2(2k+3+v)-u, \quad I[\mathcal{Y}_2] = 2k+3+2v-u,
\end{equation}
where $I[f]$ denotes the power of $\varphi$ of the function $f$ near $\varphi=0$. In this context, to respect QI to the level allowed by the breaking of QI at first order, it is reasonable to require $I[\Delta B_{2c}^\mathrm{QI}]\geq\min_{n\in\{0,1,2\}}\{I[\mathcal{Y}_n]\}$, because otherwise the breaking at second order would be larger than that inherited from the first order violation of QI in the neighbourhood of $\varphi = 0$. 
\par
To proceed further, we must say something regarding the indices $u,~v$ and $k$. The constraint of omnigeneity at the bottom of the well imposes severe constraints in $u$ and $v$ \citep{rodriguez2023higher}. If $u$ is very large, and thus the leading order field is very flat near the bottom of the well, the first order perturbation can easily lead to the appearance of defects in the magnetic field. I.e. a topological change in the contours near the bottom, which were called \textit{puddles} in \cite{rodriguez2023higher}. This immediately leads to de-confinement of deeply trapped particles, and thus the requirement of omnigeneity at first order requires $2v\geq u$, and only equal in the special case $(u,v)=(2,1)$. With this constraint in mind, $I[\mathcal{Y}_1]\geq I[\alpha_\mathrm{buf}^2]$ and $I[\mathcal{Y}_2]\geq I[\alpha_\mathrm{buf}]$, so that the contribution of $\mathcal{Y}_1$ may be neglected close to the bottom of the well. Thus, we are left with $I[\Delta B_{2c}^\mathrm{QI}]\geq I[\alpha_\mathrm{buf}]+\min\{v,2v-u\}$, meaning that, at least, it is reasonable to impose the second order QI condition in the region where the buffer is weak. 
\par 
A more quantitative measure of the extent of the region near the bottom can be obtained by computing $\mathcal{Y}_n$ functions explicitly. Note, though, that an interpretation of what a `reasonable' absolute value of this term is will depend on the precise features about an omnigeneous field that one seeks. Different considerations (say neoclassical transport through $\epsilon_\mathrm{eff}$ \citep{nemov1999evaluation}, fast particle confinement \citep{nemov2008poloidal,velasco2021model}, or other derived QI properties such as vanishing bootstrap current \citep{Helander_2009}) will lead to different measures. We leave that for development in future applications and satisfy ourselves with the simple criterion of satisfying QI in the region where the buffer function is small.

\section{Explicit form of second order equations for $B_2$} \label{app:2nd_order_eqns}
In this appendix we reproduce the key equations relating the harmonic components of $B_2$ and $X_2$ for reference. We do so instead of simply referring to \cite{landreman2019} (LS) due to the important role played by these equations. Reading off Eqs.~(A35) and (A36),
\begin{subequations}
    \begin{equation}
        B_{2c} = \kappa B_0X_{2c}-\hat{\mathcal{T}}_c+\frac{(B_{1c}^\mathrm{QI})^2}{2B_0}\cos2\alpha_1, \tag{\ref{eqn:X2c}}
    \end{equation}
    \begin{equation}
        B_{2s}=\kappa B_0X_{2s}-\hat{\mathcal{T}}_s+\frac{(B_{1c}^\mathrm{QI})^2}{2B_0}\sin2\alpha_1. \tag{\ref{eqn:X2s}}
    \end{equation} 
\end{subequations}
where
\begin{subequations}
    \begin{gather}
        B_{1c}^\mathrm{QI}=B_0\bar{d}\kappa, \\
        \hat{\mathcal{T}}_c=\frac{B_0}{\ell'}\left[Z_{2c}'+2\bar{\iota}_0 Z_{2s}+\frac{q_c^2-q_s^2+r_c^2-r_s^2}{4\ell'}\right], \\
        \hat{\mathcal{T}}_s=\frac{B_0}{\ell'}\left[Z_{2s}'-2\bar{\iota}_0 Z_{2c}+\frac{q_cq_s+r_cr_s}{2\ell'}\right].
    \end{gather}
\end{subequations}
and using the notation of Eqs.~(A37)-(A40) in LS,
\begin{subequations}
    \begin{gather}
        q_s=X_{1s}'-\bar{\iota}_0X_{1c}-\tau\ell'Y_{1s}, \\
        q_c=X_{1c}'+\bar{\iota}_0X_{1s}-\tau\ell'Y_{1c}, \\
        r_s=Y_{1s}'-\bar{\iota}_0Y_{1c}+\tau\ell'X_{1s}, \\
        r_c=Y_{1c}'+\bar{\iota}_0Y_{1s}+\tau\ell'X_{1c}. 
    \end{gather}
\end{subequations}
The parity of the different terms can be easily traced down by using the information regarding the parity of first and second order near-axis functions in Tables~\ref{tab:firs_order_parity} and \ref{tab:second_order_parity}, in addition to $\kappa$ and $\tau$ necessarily being odd and even about the minimum of the well.

\section{Alternative construction of 2nd order omnigeneous field} \label{app:alt_2nd_omni}
In the main text, Section~\ref{sec:2nd_order_qi_conditions}, we presented a way of constructing an equilibrium field to 2nd order that satisfied omnigeneity to 2nd order. This \textit{minimal shaping} approach, in an attempt to minimise second order shaping, considered the simple choice of $X_{2c}=0=X_{2s}$, and focused on optimising the 0th and 1st order near-axis choices to satisfy the QI constraints. 
\par
There is naturally an alternative approach, in which the role of the first order degrees of freedom is minimised, and the second order shaping exploited as much as possible to satisfy the omnigeneous considerations on $|\mathbf{B}|$. We call this the \textit{omnigeneous completion} of the field. To proceed this way, one must follow the following steps.
\begin{enumerate}
    \item \underline{Consistency condition}: the first step in the construction is to make sure that the constraints at flattening points are satisfied. For a zero of curvature of order $v$ (that is, $\kappa\sim\phi^v$ locally),
    \begin{equation}
        \frac{\mathrm{d}^{l}}{\mathrm{d}\varphi^l}\left[\hat{\mathcal{T}}_c \cos{2\alpha_1}+\hat{\mathcal{T}}_s \sin{2\alpha_1}+\frac{B_0^2}{4}\left(\frac{d^2}{B_0'}\right)'\right]_{\varphi=\varphi_c}=0,
    \end{equation}
    for $l\in\mathbb{N}^\mathrm{even}\cup \{0\}$ such that $0\leq l<v$ where $\varphi_c$ represents the flattening point.\footnote{Note that a set of constraints like this arise from every flattening point in the axis. If we specialise to a single magnetic well per field period, then there are in principle two points worth considering, namely, the bottom and top of the well. However, in practice, we may ignore the behaviour near the top of the well, where the first order field cannot be omnigeneous. In fact, one should think of the construction as a way of building the second order field in a finite region about the bottom of the well where omnigeneity at second order is of interest.}  These are $(v+1)/2$ (for $v$ odd) constraints on lower order quantities, which may be thought of as constraints on the local values of $\Bar{d}^{(l+2)}$. This is the only step in which the construction restricts the lower order near-axis choices. One must find a first order construction that satisfies these constraints.
    
    \item \underline{Choose the second order shaping:} once the conditions near the flattening points are satisfied, one can then solve Eq.~(\ref{eqn:B2c_qi_eq}) for $\Tilde{X}_{2c}$ to find a shaping that enforces the omnigeneous behaviour everywhere. The key observation is that one may always do so provided step one is taken first; otherwise, the construction of $\tilde{X}_{2c}$ is not well-posed. After choosing $\Tilde{X}_{2c}$ accordingly, the problem still retains some freedom in the choice of $\Tilde{X}_{2s}$. Besides it being even to preserve stellarator symmetry, it is a free function. 
    
    \item \underline{Construct second order field}: with these we have all the elements needed to revert Eqs.~(\ref{eqn:X2c})-(\ref{eqn:X2s}), and complete the second order construction (Figure~\ref{fig:2nd_order_nae_equilibrium}).
\end{enumerate}

The little control on the second order shaping of this approach leads, generally and in practice, to near-axis equilibria with large amounts of shaping, which limits the range of validity of the near-axis description itself. This break-down manifests in the form of nonphysically intersecting flux surfaces \citep{landreman2021a}. Thus we so far generally favour the minimal shaping alternative presented in the main text.

\bibliographystyle{jpp}

\bibliography{jpp-instructions}

\end{document}